%% file: main.tex
\documentclass[%
 preprint,
 amsmath,amssymb,
aps,
]{revtex4-1}

\usepackage{graphicx}
\usepackage{dcolumn}
\usepackage{bm}

\usepackage{color}
\usepackage{float}
\usepackage[letterpaper, portrait, margin=1in]{geometry}
\usepackage[utf8]{inputenc}
\usepackage{graphicx}
\usepackage{subcaption}
\usepackage{multirow}
\usepackage{bibunits}
\graphicspath{{Images_new/}}
\begin{document}

\preprint{APS/123-QED}

\title{Thermodynamic Stabilization of Precipitates through Interface Segregation: Chemical Effects}

\author{Sourabh B Kadambi}
\author{Srikanth Patala}%
 \email{spatala@ncsu.edu}
\affiliation{%
 Department of Materials Science and Engineering \\
 North Carolina State University
}%

\begin{abstract}
  Precipitation hardening, which relies on a high density of
  intermetallic precipitates, is a commonly utilized technique for
  strengthening structural alloys.  At high temperatures, however, the
  precipitates coarsen to reduce the excess energy of the interface,
  resulting in a significant reduction in the strengthening provided
  by the precipitates. In certain ternary alloys, the secondary solute
  segregates to the interface and results in the formation of a high
  density of nanosized precipitates that provide enhanced strength and
  are resistant to coarsening. To understand the chemical effects
  involved, and to identify such segregating systems, we develop a
  thermodynamic model using the framework of the regular
  nanocrystalline solution model. For various global compositions,
  temperatures and thermodynamic parameters, equilibrium configuration
  of Mg-Sn-Zn alloy is evaluated by minimizing the Gibbs free energy
  function with respect to the region-specific (bulk solid-solution,
  interface and precipitate) concentrations and sizes. The results
  show that Mg$_2$Sn precipitates can be stabilized to nanoscale sizes
  through Zn segregation to Mg/Mg$_2$Sn interface, and the
  precipitates can be stabilized against coarsening at
  high-temperatures by providing a larger Zn concentration in the
  system. Together with the inclusion of elastic strain energy effects
  and the input of computationally informed interface thermodynamic
  parameters in the future, the model is expected to provide a more realistic
  prediction of segregation and precipitate stabilization in ternary
  alloys of structural importance.

\begin{description}
\item[Keywords] Statistical thermodynamics, Intermetallic precipitatation, Coarsening, Solute segregation, Heterophase interface

\end{description}
\end{abstract}

\pacs{Valid PACS appear here}
\maketitle


\begin{bibunit}[apsrev4-1]
\input{Introduction}

\input{Methodology}

\input{ParametricStudy}

\input{Conclusions}

\section{ACKNOWLEDGMENTS}
This work was primarily supported by the U.S. National Science Foundation under Grant No. DMR-1554270, with partial support from North Carolina State University.

\newpage

\putbib


\end{bibunit}

\clearpage

\input{SupportingInfo.tex}
\end{document}

%% file: Introduction.tex
\section{Introduction}
\label{sec:Intro}

In structural alloys, one of the most commonly utilized strengthening
mechanism involves the precipitation of ordered intermetallic
compounds. For example, the precipitation of L1$_2$ Al$_3$Sc
\cite{hyland1992homogeneous,royset2013scandium} and Al$_3$Li
\cite{gregson1985microstructural,flower1987solid,noble1997microstructural}
intermetallics in aluminum has led to significant improvements in
yield strength. Similar effects were observed in certain maraging
\cite{floreen1968physical,vasudevan1990precipitation} and stainless
steels \cite{judy1973properties}. The beneficial properties of intermetallic precipitation
are not just confined to fcc alloy systems but have also been
observed in hexagonal alloys of titanium
\cite{madsen1995separating,ramos2006peritectoid} and, more recently,
in magnesium alloys \cite{antion2003hardening, apps2003precipitation,
  mendis2006refinement}. Even in iron-based bcc alloys, L2$_1$
precipitates have resulted in significant improvements in yield
strength \cite{jack1975low, dimiduk1988structural,
  perrier2011precipitation}.

Precipitation hardening is usually attributed to two
primary mechanisms - (i) dislocation cutting-through or bowing-around
dispersed precipitate particles (dispersion hardening
\cite{hazzledine1974coplanar}) and (ii) dislocations interacting with
the coherency strains of precipitates (coherency strain hardening
\cite{mott1940attempt, gladman1999precipitation}).

In addition to mechanical properties, thermal stability of certain fcc
alloys can be dramatically improved through the precipitation of
ordered intermetallics. For example, precipitation of cubic L1$_2$
intermetallics, in nickel \cite{kear1970stacking, timmins1986negative,
  condat1987shearing, sarosi2005imaging, sarosi2006direct} and
cobalt-based alloys \cite{chinen2007new,suzuki2007flow}, results in an
improved high-temperature yield strength
\cite{pope1984mechanical}. While the improved mechanical properties
have a non-trivial dependence on the size, density and crystallography
of the precipitates, it is generally true that \emph{a higher density
  of smaller precipitates that are resistant to coarsening} improves
both the strengthening characteristics and high-temperature
stability. For example, while the separate addition of Zr or Sc to
Aluminum alloys results in increased tensile strength and resistance
to recrystallization by forming ordered precipitates, the combined
effect is considerably larger \cite{yelagin1985influence,
  davydov1996alloying, toropova1998advanced, fuller2003mechanical,
  riddle2004study, fuller2005temporal1, fuller2005temporal2}. This is
due to the formation of a higher density of small precipitates driven
by Zr segregation to Al/Al$_3$Sc hetero-phase interfaces
\cite{clouet2006complex}.

Zr segregation in Al-Sc-Zr alloys has been attributed primarily to
kinetic effects - the low diffusion coefficient of Zr compared to
Sc restricts the partitioning of Zr to just the interface layer rather
than the core of Al$_3$Sc \cite{clouet2006complex}. However, in
general, both kinetic and thermodynamic factors may promote solute
segregation to an interface. Some observed examples include:
segregation of Mg \cite{marquis2005coarsening}, Zr
\cite{tolley2005segregation} and Mg+Ag \cite{kang2014determination} to
$\alpha$-Al/Al$_3$Sc interface; Sc \cite{yang2016influence}, Si
\cite{biswas2011precipitates}, Ag \cite{rosalie2012silver} and Si+Mg \cite{biswas2010simultaneous}
to $\alpha$-Al/Al$_2$Cu interface; Gd+Zn \cite{nie2008solute} to
Mg/Mg$_5$Zn interface; Zn \cite{liu2016interphase,liu2017zn} to Mg/Mg$_2$Sn
interface. The thermodynamic driving force for solute segregation to
the interface is attributed to a reduction in the interfacial
free-energy \cite{weissmuller1993alloy, weissmuller1994alloy,
  kirchheim2002grain, wynblatt2006anisotropy, chookajorn2012design,
  murdoch2013stability, murdoch2013estimation}, which is attributed primarily to two factors:
(a) favorable chemical interactions of solute element at the interface
over that in the bulk; and (b) reduction of solute size misfit strain energy
\cite{voorhees2006alloys}. The reduction in the interfacial
free-energy $\gamma$ further results in a decrease in the coarsening
kinetics at elevated temperatures \cite{kuehmann1996ostwald}.


The afore-mentioned examples are but a few among a large number of
binary alloy systems that favor intermetallic precipitation. For
example, other commonly observed binary intermetallics in structural
alloys include: Cu$_3$Al, Cu$_3$Sn, Cu$_3$Ti, Cu$_3$Au, Ni$_3$Al,
Ni$_3$Ti, Ni$_3$Nb, Ni$_3$Si, MgZn, Mg$_2$Si, Mg$_3$Nd, Mg$_2$Cu,
Mg$_2$Ni. By carefully introducing ternary atoms that will segregate
to the the precipitate/matrix interface, it will be possible to
stabilize much smaller precipitates that are resistant to coarsening
at high-temperatures.
To identify such systems, in section \ref{sec:Atreat}, we develop a
thermodynamic model that describes the energetics of simple ternary
alloys (A-B-C), where the binary A-B system favors the precipitation
of the ordered compound $A_{m}B_{n}$ and the impurity C atoms may
segregate to the interface between the matrix and the $A_{m}B_{n}$
precipitate.

It is assumed that the segregation is promoted just through favorable
chemical interactions at the interface. The important contribution of
elastic strain is ignored, for now, so that we can build a simple
thermodynamic model for the ternary alloy system. Future work will
focus on incorporating the strain energy effects. Under these
simplifying conditions, the model is assumed to be applicable to
precipitating systems with incoherent interfaces (e.g. incoherent
equilibrium $\theta$ phase in Al-Cu alloys
\cite{biswas2011precipitates}). In this article, we present the
thermodynamic model using the ternary Mg-Sn-Zn alloy system, where
recent experimental studies \cite{liu2016interphase,liu2017zn} have
shown the segregation of Zn to all Mg/Mg$_2$Sn interfaces irrespective
of interface structure and orientation relationship and that the
segregation is not limited by the kinetics of Zn diffusion.  This
observation suggests the presence of a chemical driving force for
heterophase interface segregation. In section \ref{par_study}, the
variations in equilibrium precipitate sizes and the Gibbs interfacial
excess of the solute atoms, for the Mg-Sn-Zn ternary alloy, as a
function of different interaction energy parameters and temperature
are presented.

%% file: Methodology.tex
\section{ANALYTICAL TREATMENT}
\label{sec:Atreat}

In this section, we present a thermodynamic model to study solute
segregation to the interface between a solid-solution bulk phase and a
binary ordered-intermetallic-compound. The Gibbs free energies of the bulk
and the interface regions of the system are described using a
statistical-thermodynamic framework involving the regular solution
assumptions of random mixing and nearest-neighbor pairwise interaction
\cite{dehoff2006thermodynamics}. This follows the regular
nanocrystalline solution model developed by Trelewicz and Schuh
\citep{trelewicz2009grain} and extended to ternary systems by Saber et
al. \cite{saber2013thermodynamic}. Energy contribution of the
precipitate region is described using the Gibbs free energy of
formation of the intermetallic compound; this avoids the complexity
involved in sublattice models. The model for a binary alloy system is
presented first. The binary system consists of a solvent element, $A$,
and a solute element, $B$, that favor precipitation of an
intermetallic compound of the type $ A_{m}B_{n} $ from
solid-solution. The model is subsequently extended to a ternary alloy
system, wherein a secondary solute element, $C$, is considered to be
soluble in the bulk and the interface solid-solutions but assumed to
be insoluble in the precipitate. The bulk and interface regions are
provided with distinct descriptions of energetic parameters and
compositions. This feature allows equilibrium segregation of solute
to the interface for energetic parameters favoring the reduction in
interfacial energy, and thus in the system free energy, on solute
segregation. The free energy function, important relations and
definitions pertaining to the model are presented below; the complete
derivation is given in section \ref{SI1:Atreat} of Supplemental
Materials.

\subsection{Binary Model} \label{method:binary}

The alloy system, consisting of a heterogeneous distribution of $A$
and $B$ atoms, is divided into three distinct regions---the bulk, the
precipitate and the bulk/precipitate interface regions, denoted by
$b$, $p$ and $i$, respectively.  The global concentration of $B$ in
the system, $x_{\circ}$, can be expressed as a function of region-specific
concentrations, $x_b$, $x_i$ and $x_p$, and volume fractions, $f_b$,
$f_i$ and $f_p$, using the mass balance relation as:

\begin{equation} \label{eqm:cons_x}
  x_{\circ} = x_b \left(1-f_i-f_p \right) + x_i f_i + x_p f_p. \\
\end{equation}

\noindent Here, $f_b = 1-f_i-f_p$, and since the precipitate is an
intermetallic compound, concentration in $p$ is stoichiometric with
$x_p = n/m$. For a closed system of given $x_{\circ}$, this relation imposes
constraint on the values that the variables $x_b$, $x_i$, $f_i$ and
$f_p$ can take simultaneously.

The assumptions of random mixing and nearest-neighbor pairwise
interaction between the atoms of bulk solid-solution result in the
following expression for the Gibbs free energy of mixing of $b$,
$\Delta \bar{G}_{b}^{\,mix}$:

\begin{align} \label{eq:dGb_bin}
  \begin{split}
	\Delta \bar{G}_{b}^{\,mix} =& \left\{\omega_b^{AB} \left(1 - x_b \right) x_b z_b \right. \\
    &+ \left. RT \left[\left(1-x_b \right) \text{ln} \left(1-x_b \right) + x_b\,\text{ln}\,x_b \right]\right\} \left(1-f_{i}-f_p\right),
    \end{split}
\end{align}

\noindent where, $z_b$ is the coordination number and and
$\omega_b^{AB}$ is the regular solution interaction parameter,
specific to region $b$. The bar over $\Delta \bar{G}_b^{mix}$ defines
the quantity per mole of atoms in the system and applies to all
quantities represented this way in this paper; $\omega_b^{AB}$ is
defined per mole of atoms in $b$.  This expression is essentially the
free energy of mixing obtained in the regular solution model, scaled
by the size of the bulk region relative to the system (i.e.
$f_b = 1-f_{i}-f_p$). The first term in the expression corresponds to
the excess enthalpy of mixing, obtained from the internal energy of
mixing assuming negligible volume change during mixing. Here,
$\omega_b^{AB}$ accounts for the energy difference involved in the
formation of unlike $AB$ bonds from like $AA$ and $BB$ bonds having
energies characteristic of the bulk region; this is given by:

\begin{align} \label{eq:omb_ab} 
    \omega_b^{AB} &= E_b^{AB} - \frac{E_b^{AA}+E_b^{BB}}{2}.
\end{align}

\noindent The part of term multiplied to $\omega_b^{AB}$, along with
the bulk volume fraction $f_b$, represents the number of $AB$ bonds,
$N_b^{AB}$, in the bulk region. The first term thus corresponds to the
enthalpy of forming unlike bonds from like bonds. The second term in
Eq. (\ref{eq:dGb_bin}) represents the ideal entropy of mixing in $b$.

In general, the number of bonds of type $kl$ in region $r$,
$N_r^{kl}$, is obtained as:

\begin{equation} \label{eq:bond_cons1m} 
	N_r^{kl} = N_r^{bonds}P_r^{kl},
\end{equation}

\noindent where, $r$ refers to the bonding regions of $b$ and $i$, and
the transition regions $ib$ and $ip$, occurring between atoms of $i$
and $b$, and atoms of $i$ and $p$, respectively. $N_r^{bonds}$ is the
total number of bonds in $r$; $P_r^{kl}$ is the statistical
probability that the bond in $r$ is of type $kl$, and is obtained as a
function of the composition in each region $r$. For the binary system
with a total number of $N_{\circ}$ atoms of $A$ and $B$, expressions for
various $N_r^{bonds}$ and $P_r^{kl}$ are presented in Table
\ref{table:binary_main}.

\begin{table}[H]
\begin{center}
  \caption{Number of bonds, bond probabilities and bond energies of
    various bond-types specific to the different bonding-regions in
    the binary system.}
  \begin{tabular}{c c c c c}
    \hline \hline
    \multirow{3}{6em}{\centering Region} & \multirow{3}{8.5em}{\centering Number of bonds in region $r$, $ N_r^{bonds} $} & \multirow{3}{4em}{\centering Bond type, $kl$} & \multirow{3}{4em}{\centering Bond energy} & \multirow{3}{10em}{\centering Bond probability, $ P_r^{kl} $} \\
                                         & & & & \\
                                         & & & & \\ [0.5ex]
    \hline
                                         & & & & \\
    \multirow{1}{6em}{\centering Bulk, $b$} & \multirow{1}{8.5em}{\centering $ N_{\circ}\left(1-f_i-f_p\right)\frac{z_b}{2} $} & AB & $E_b^{AB}$ & $2\left(1-x_b\right)x_b$ \\
    [0.3ex] \\
    \multirow{3}{6em}{\centering Interface, $i$} & \multirow{3}{8.5em}{\centering $ N_{\circ} f_i \frac{z_{ii}}{2} $} & AA & $E_i^{AA}$ & $\left(1-x_i\right)^2$ \\
                                         & & BB & $E_i^{BB}$ & $x_i^2$ \\
                                         & & AB & $E_i^{AB}$ & $2\left(1-x_i\right)x_i$ \\
    [0.3ex] \\
    \multirow{3}{6em}{\centering Interface-bulk transition, $ib$} & \multirow{3}{8.5em}{\centering $ N_{\circ} f_i \frac{z_{ib}}{2} $} & AA & $E_i^{AA}$ & $\left(1-x_i\right)\left(1-x_b\right)$ \\
                                         & & BB & $E_i^{BB}$ & $x_i x_b$ \\
                                         & & AB & $E_i^{AB}$ & $\left(1-x_i\right)x_b + \left(1-x_b\right)x_i$ \\
    [0.3ex] \\
    \multirow{3}{6em}{\centering Interface-precipitate transition, $ip$} & \multirow{3}{8.5em}{\centering $ N_{\circ} f_i \frac{z_{ip}}{2} $} & AA & $E_i^{AA}$ & $\left(1-x_i\right)\left(1-x_p^{i/f}\right)$ \\
                                         & & BB & $E_i^{BB}$ & $x_i x_p^{i/f}$ \\
                                         & & AB & $E_i^{AB}$ & $\left(1-x_i\right)x_p^{i/f} + \left(1-x_p^{i/f}\right)x_i$ \\ [2ex]
    \hline \hline
  \end{tabular}
  \label{table:binary_main}
\end{center}
\end{table}

The Gibbs free energy of mixing for the formation of interface
solid-solution, $\Delta \bar{G}_i^{\,mix}$, is obtained similar to $b$
following the regular solution model as:

\begin{align} \label{eq:dGi_bin}
  \begin{split}
	\Delta \bar{G}_{i}^{\,mix} =& \left\{ \omega_{i}^{AB} \left(1 - x_{i} \right) x_{i} z_{ii}\right. + \left[\delta_i^{AA}\left(1-x_{i} \right) + \delta_i^{BB}x_{i} \right]\frac{z_{ii}}{2} \\
    &+ \left. RT \left[\left(1-x_{i} \right) \text{ln}\left(1-x_{i}
      \right) + x_{i}\,\text{ln}\,x_{i} \right]\right\} f_{i}.
    \end{split}
\end{align}

\noindent Here, the interface interaction parameter, $\omega_i^{AB}$,
captures the energetics of formation of unlike bonds in region $i$
from like bonds in pure element states characteristic of $i$, and is
given by:

\begin{align} \label{eq:omi_ab} 
    \omega_i^{AB} &= E_i^{AB} - \frac{E_i^{AA}+E_i^{BB}}{2}.
\end{align}

\noindent The formation of the interface solid-solution is considered
from the same reference state as that for the formation of the bulk;
i.e. the reference state is pure $A$ and pure $B$ with bond-energies
and bond-coordination characteristic of $b$. Therefore, in addition to
the enthalpy of mixing (first term), an additional term (second term)
arises---this corresponds to the energy associated with a change of
state of components $A$ and $B$ from a state characteristic of $b$ to
one characteristic of $i$. This change of state is captured by the
parameters $\delta^{AA}_i$ and $\delta^{BB}_i$, and is defined per
mole of $A$ and $B$, respectively, as,

\begin{align} \label{eq:deli_aa} 
	\delta_i^{AA} = E_i^{AA} - \frac{z_b}{z_i} E_b^{AA} \quad
  \text{and} \quad \delta_i^{BB} = E_i^{BB} - \frac{z_b}{z_i} E_b^{BB}.
\end{align}

\noindent These parameters represent the difference between like bond
energies characteristic of the interface and the bulk and can be
directly related to the interface free energies of $A$ and $B$ (see
Supporting Materials \ref{SI3:delta}); $z_b/z_i$ accounts for the
different atomic coordination of $b$ and $i$. Analogous to
Eq. \ref{eq:dGb_bin}, the terms multiplying the energy parameters in
Eq. \ref{eq:dGi_bin} are obtained form Table
\ref{table:binary_main}. The term coupled to $\omega_i^{AB}$ is
$N_i^{AB}$; and terms coupled to $\delta_i^{AA}$
(i.e. $f_i(1-x_i)z_{ii}/2$) and $\delta_i^{BB}$
(i.e. $f_ix_iz_{ii}/2$) represent the number of $AA$ bonds
($ N_i^{AA} + N_i^{AB}/2$) and $BB$ bonds
($ N_i^{BB} + N_i^{AB}/2$), respectively, in the pure element
reference state of $i$. The terms coupled to the parameters
$\omega_b$, $\omega_i$ and $\delta_i$ in subsequent equations can be
interpreted in a similar manner.

The interface region is considered as one layer of atoms
between the bulk and the precipitate regions. Thus, of the total
coordination $z_i$ per interface atom, $z_{ii}$ connects to
neighboring interface atoms, while $z_{ib}$ and $z_{ip}$ connect to
$b$ and $p$ atoms, respectively, that lie adjacent to region $i$. To
account for the non-random distribution of atoms in $p$ and the
dependence on the interface plane, concentration at the atomic layer
of precipitate adjacent to $i$ is defined distinctly as
$x_p^{i/f}$. Random distribution of atoms with a concentration of
$x_b$ is assumed to be applicable to the atomic layer of $b$ adjacent
to $i$. $ib$ and $ip$ bonds are defined to have bond-energies
characteristic of $i$. Following these considerations, the Gibbs free energy for the formation of the
transition regions $ib$ and $ip$, $\Delta \bar{G}_{ib}$ and
$\Delta \bar{G}_{ip}$, respectively, are obtained as:

\begin{align} \label{eq:dG_ib_binary}
  \begin{split}
	\Delta \bar{G}_{ib} =&\; \omega_{i}^{AB} \left[ \left(1-x_b \right) x_{i}  + \left(1-x_i \right) x_b \right] f_{i} \frac{z_{ib}}{2} \\
    &+ \left[\delta_i^{AA}\left(1-x_{i} + 1-x_{b} \right) + \delta_i^{BB}\left(x_{i}+x_b \right)  \right]f_{i} \frac{z_{ib}}{4},
    \end{split}
\end{align}

\begin{align} \label{eq:dG_ip_binary}
  \begin{split}
	\Delta \bar{G}_{ip} =&\; \omega_{i}^{AB} \left[ \left(1-x_p^{i/f} \right)x_{i}  + \left(1-x_{i} \right)x_p^{i/f}  \right] f_{i} \frac{z_{ip}}{2} \\
    & {}+ \left[\delta_i^{AA}\left(1-x_{i} + 1-x_{p}^{i/f} \right) + \delta_i^{BB}\left(x_{i}+x_p^{i/f} \right) \right]f_{i} \frac{z_{ip}}{4}.
    \end{split}
\end{align}

\noindent Analogous to the interpretation of Eq. \ref{eq:dGi_bin}, the
first terms of Eqs. \ref{eq:dG_ib_binary} and \ref{eq:dG_ip_binary}
represent the enthalpy change associated with the formation of unlike
bonds of the corresponding transition regions from like bonds of pure
$A$ and pure $B$ states characteristic of $i$. The second terms
corresponds to the additional enthalpy associated with the change of
state of the pure elements from energies characteristic of $b$ to $i$.
 
While the free energy expressions presented in Eqs. \ref{eq:dGb_bin},
\ref{eq:dGi_bin}, \ref{eq:dG_ib_binary} and \ref{eq:dG_ip_binary}
considered the reference states as pure $A$ and pure $B$ with crystal
structures characteristic of the bulk solid-solution, $A$ or $B$ or
both, however, may have a crystal structure preference different from
the bulk. To enable treatment of such systems, the initial or standard
state is chosen as pure $A$ and pure $B$ at $T$, having crystal
structures specific to their Standard Element Reference State
(i.e. the most stable state of an element at $298.15$ K and $10^5$
Pa \cite{dinsdale1991sgte}). The molar free energies involved in the conversion of $A$ and $B$
from the standard state at $T$ to the reference state characteristic
of $b$, at the same $T$, are defined as $\Delta \bar{G}_{ref(b)}^A$
and $\Delta \bar{G}_{ref(b)}^B$, respectively. The total free energy
change associated with this change in reference state,
$\Delta \bar{G}_{b,i}^{\,ref}$, for atoms forming $b$ and $i$ regions
of the system is obtained as a mole fraction-weighted average of
$\Delta \bar{G}_{ref(b)}^A$ and $\Delta \bar{G}_{ref(b)}^B$, scaled by
the region-sizes as:

\begin{align} \label{eq:dG_ref_bin}
	\begin{split}
	\Delta \bar{G}^{\,ref}_{b,i} =& \left[\left(1-x_b \right) \Delta \bar{G}_{ref(b)}^A + x_b\Delta \bar{G}_{ref(b)}^B \right] \left(1 - f_{i} - f_p \right) \\
    &+ \left[\left(1-x_{i} \right) \Delta \bar{G}_{ref(b)}^A + x_{i}\Delta \bar{G}_{ref(b)}^B \right]f_{i}.
	\end{split}
\end{align} 

The Gibbs free energy for the formation of precipitate region,
$\Delta \bar{G}_p$, is obtained in terms of the molar Gibbs free
energy for the formation of $A_mB_n$ intermetallic,
$\Delta \bar{G}_f^{A_mB_n}$, from the pure element standard states of
$A$ and $B$, and is scaled by the precipitate size.

\begin{equation} \label{eq:dGp_bin}
	\Delta \bar{G}_{p} = \Delta \bar{G}_f^{A_m B_n} f_p
\end{equation}

The total Gibbs free energy, $\Delta \bar{G}_{bin}$---defined for the
formation of the binary system configuration from initial pure element
standard states of $A$ and $B$---is obtained as the summation over
free energy contributions corresponding to the formation of $b$, $i$,
$ib$, $ip$ and $p$ regions of the system from Eqs. \ref{eq:dGb_bin},
\ref{eq:dGi_bin}, \ref{eq:dG_ib_binary}, \ref{eq:dG_ip_binary},
\ref{eq:dG_ref_bin} and \ref{eq:dGp_bin}; thus,

\begin{align} \label{eq:dG_sum_bin}
	\Delta \bar{G}_{bin} = \Delta \bar{G}^{\,mix}_b + \Delta \bar{G}^{\,mix}_{i} + \Delta \bar{G}_{ib} + \Delta \bar{G}_{ip} + \Delta \bar{G}^{\,ref}_{b,i} + \Delta \bar{G}_p.
\end{align}

\subsection{Ternary Model} 
\label{method:ternary}

The binary model presented in section \ref{method:binary} is now
extended to a ternary system containing an additional element in
secondary solute $C$. The ternary alloy system consists of a
heterogeneous distribution of $A$, $B$ and $C$ atoms and, as before,
the system has three distinct regions of atom occupancy, viz. $b$, $p$
and $i$, and bonding regions of $b$, $i$, $p$, $ib$ and $ip$. We
consider $C$ to form ternary solid-solution with $A$ and $B$ in the
bulk and the interface regions, while insoluble in the intermetallic
compound. In addition to the mass balance relation for solute $B$,
given by Eq. \ref{eqm:cons_x}, mass balance relation for solute $C$ is
obtained as:

\begin{align} \label{eq:cons_y}
  \begin{split}
     y_{\circ} = y_b \left(1-f_i-f_p \right) + y_i f_i.
  \end{split}
\end{align}

\noindent Each of the variables $x_i$, $x_b$, $y_i$, $y_b$, $f_i$ and
$f_p$ can take values between $0$ and $1$. For given values of $x_{\circ}$
and $y_{\circ}$, Eqs. \ref{eqm:cons_x} and \ref{eq:cons_y} impose
constraints on the values the above variables can take
simultaneously. We will consider $x_i$, $y_i$, $f_i$ and $f_p$ as the
independent variables.

The Gibbs free energy function, $\Delta \bar{G}_{tern}$, for the
formation of the ternary system from the standard pure element
states of $A$ and $B$ is obtained, similar to that for the binary
system, as the summation over free energy contributions corresponding
to the formation of $b$, $i$, $ib$, $ip$ and $p$ regions of the
system. Thus,

\begin{equation}
\Delta \bar{G}_{tern} = \Delta \bar{G}^{\,mix}_b + \Delta \bar{G}^{\,mix}_{i} + \Delta \bar{G}_{ib} + \Delta \bar{G}_{ip} + \Delta \bar{G}^{ref}_{b,i} + \Delta \bar{G}_p,
\label{eq:dG_tern_main}
\end{equation}

\noindent where,

\begin{align*}
\Delta \bar{G}^{\,mix}_b &= \left\{ \left[\omega_b^{AB} \left(1 - x_b - y_b \right) x_b + \omega_b^{BC} x_b y_b + \omega_b^{AC} \left(1 - x_b - y_b \right) y_b \right] z_b \right. \nonumber \\
 &+ \left. RT \left[\left(1-x_b-y_b \right) \text{ln} \left(1-x_b-y_b \right) + x_b\,\text{ln}\,x_b + y_b\,\text{ln}\,y_b \right] \right\} \left(1-f_{i}-f_p\right),
\end{align*}

\vspace{-30pt}

\begin{align*}
    \Delta \bar{G}^{\,mix}_{i} =& \left\{ \left[\omega_{i}^{AB} \left(1 - x_{i} - y_{i} \right) x_{i} + \omega_{i}^{BC} x_{i} y_{i} + \omega_{i}^{AC} \left(1 - x_{i} - y_{i} \right) y_{i} \right] z_{ii}\right. \nonumber \\
    &+ \left[\delta_i^{AA}\left(1-x_{i}-y_{i}\right) + \delta_i^{BB}x_{i} + \delta_i^{CC}y_{i} \right] \frac{z_{ii}}{2} \nonumber \\
    &+ \left. RT \left[\left(1-x_{i}-y_{i} \right)
      \text{ln}\left(1-x_{i}-y_{i} \right) + x_{i}\,\text{ln}\,x_{i} +
      y_{i}\,\text{ln}\,y_{i} \right] \right\} f_{i},
\end{align*}

\vspace{-30pt}

\begin{align*}
      \Delta \bar{G}_{ib} =& \left\{\omega_{i}^{AB} \left[ \left(1-x_b-y_b \right)x_{i} + \left(1-x_{i}-y_{i} \right)x_b \right] + \omega_{i}^{BC} \left(x_b y_{i} + x_{i} y_b \right) \right. \nonumber\\
    &\qquad+ \left. \omega_{i}^{AC} \left[\left(1-x_b-y_b\right)y_{i} + \left(1-x_{i}-y_{i} \right)y_b \right] \right\} f_{i} \frac{z_{ib}}{2} \nonumber\\
    &+ \left[\delta_i^{AA}\left(1-x_{i}-y_{i} + 1-x_{b}-y_{b} \right) + \delta_i^{BB}\left(x_{i}+x_b \right) + \delta_i^{CC}\left(y_i + y_b\right)  \right]f_{i} \frac{z_{ib}}{4},
\end{align*}

\vspace{-30pt}

\begin{align*}
    \Delta \bar{G}_{ip} =& \left\{\omega_{i}^{AB} \left[ \left(1-x_p^{i/f} \right)x_{i} + \left(1-x_{i}-y_{i} \right)x_p^{i/f} \right] + \omega_{i}^{BC} x_p^{i/f} y_{i} \right. \nonumber \\
    &\qquad+ \left. \omega_{i}^{AC} \left(1-x_p^{i/f} \right)y_{i} \right\} f_{i} \frac{z_{ip}}{2} \nonumber \\
    &+ \left[\delta_i^{AA}\left(1-x_{i}-y_{i} + 1-x_{p}^{i/f} \right) + \delta_i^{BB}\left(x_{i}+x_{p}^{i/f} \right) + \delta_i^{CC} y_{i} \right]f_{i} \frac{z_{ip}}{4},
\end{align*}

\vspace{-30pt}

\begin{align*}
    \Delta \bar{G}^{\,ref}_{b,i} =& \left[\left(1-x_b-y_b\right) \Delta G_{ref(b)}^A + x_b\Delta G_{ref(b)}^B + y_b\Delta G_{ref(b)}^C \right] \left(1 - f_{i} - f_p \right) \nonumber\\
    &+ \left[\left(1-x_{i}-y_{i}\right) \Delta G_{ref(b)}^A + x_{i}\Delta G_{ref(b)}^B + y_{i}\Delta G_{ref(b)}^C \right]f_{i},
\end{align*}

\vspace{-30pt}
\begin{align*}
    \Delta \bar{G}_{p} =& \Delta \bar{G}_f^{A_m B_n} f_p.
\end{align*}

\noindent In the above equations, the interaction parameter for
different regions is given by
$\omega_r^{kl} = E_r^{kl} - \frac{E_r^{kk}+E_i^{ll}}{2}$, where $r$
refers to $b$ or $i$, $kl$ refers to unlike bonds $AB$, $BC$ or $AC$,
and $kk$ or $ll$ refer to like bonds $AA$, $BB$ or $CC$. The energy
penalty arising from the difference in bond energies of like bonds
characteristic of the interface and that of the bulk is
$\delta_i^{kk} = E_i^{kk} - \frac{z_b}{z_i} E_b^{kk}$. The terms
multiplying $\omega_b$, $\omega_i$ and $\delta_i$ (including the
volume fraction and the coordination number) correspond to the number
of bonds of the type defined by $\omega_b$, $\omega_i$ or $\delta_i$,
and were obtained, as described in section \ref{method:binary}, using
Eq. \ref{eq:bond_cons1m}; the number of bonds specific to each
bonding region, and the bond probabilities of the various bond types
are presented in Table \ref{table:ternary}. The
$\Delta \bar{G}_{tern}$ function given Eq. \ref{eq:dG_tern_main} is
the main result of our model.

\begin{table}[H]
\begin{center}
  \caption{Number of bonds, bond probabilities and bond energies of
    various bond-types specific to the different bonding-regions in
    the ternary system.}
  \begin{tabular}{c c c c c}
    \hline \hline
    \multirow{3}{6em}{\centering Region} & \multirow{3}{8.5em}{\centering Number of bonds in region $r$, $ N_r^{bonds} $} & \multirow{3}{4em}{\centering Bond type, $kl$} & \multirow{3}{4em}{\centering Bond energy} & \multirow{3}{10em}{\centering Bond probability, $ P_r^{kl} $} \\
                                         & & & & \\
                                         & & & & \\ [0.5ex]
    \hline
                                         & & & & \\
    \multirow{3}{6em}{\centering Bulk, $b$} & \multirow{3}{8.5em}{\centering $ N_{\circ}\left(1-f_i-f_p\right)\frac{z_b}{2} $} & AB & $E_b^{AB}$ & $2\left(1-x_b-y_b\right)x_b$ \\
                                         & & BC & $E_b^{BC}$ & $2x_by_b$ \\
                                         & & AC & $E_b^{AC}$ & $2\left(1-x_b-y_b\right)y_b$ \\
    [0.3ex] \\
    \multirow{6}{6em}{\centering Interface, $i$} & \multirow{6}{8.5em}{\centering $ N_{\circ} f_i \frac{z_{ii}}{2} $} & AA & $E_i^{AA}$ & $\left(1-x_i-y_i\right)^2$ \\
                                         & & BB & $E_i^{BB}$ & $x_i^2$ \\
                                         & & CC & $E_i^{CC}$ & $y_i^2$ \\
                                         & & AB & $E_i^{AB}$ & $2\left(1-x_i-y_i\right)x_i$ \\
                                         & & BC & $E_i^{BC}$ & $2x_iy_i$ \\
                                         & & AC & $E_i^{AC}$ & $2\left(1-x_i-y_i\right)y_i$ \\
    [0.3ex] \\
    \multirow{6}{6em}{\centering Interface-bulk transition, $ib$} & \multirow{6}{8.5em}{\centering $ N_{\circ} f_i \frac{z_{ib}}{2} $} & AA & $E_i^{AA}$ & $\left(1-x_i-y_i\right)\left(1-x_b-y_b\right)$ \\
                                         & & BB & $E_i^{BB}$ & $x_i x_b$ \\
                                         & & CC & $E_i^{CC}$ & $y_i y_b$ \\
                                         & & AB & $E_i^{AB}$ & $\left(1-x_i-y_i\right)x_b + \left(1-x_b-y_b\right)x_i$ \\
                                         & & BC & $E_i^{BC}$ & $x_i y_b + x_by_i$ \\
                                         & & AC & $E_i^{AC}$ & $\left(1-x_i-y_i\right)y_b + \left(1-x_b-y_b\right)y_i$ \\
    [0.3ex] \\
    \multirow{5}{6em}{\centering Interface-precipitate transition, $ip$} & \multirow{5}{8.5em}{\centering $ N_{\circ} f_i \frac{z_{ip}}{2} $} & AA & $E_i^{AA}$ & $\left(1-x_i-y_i\right)\left(1-x_p^{if}\right)$ \\
                                         & & BB & $E_i^{BB}$ & $x_i x_p^{i/f}$ \\
                                         & & AB & $E_i^{AB}$ & $\left(1-x_i-y_i\right)x_p^{i/f} + \left(1-x_p^{i/f}\right)x_i$ \\
                                         & & BC & $E_i^{BC}$ & $y_ix_p^{i/f}$ \\
                                         & & AC & $E_i^{AC}$ & $y_i\left(1-x_p^{i/f}\right)$ \\ [2ex]
    \hline \hline
  \end{tabular}
  \label{table:ternary}
\end{center}
\end{table}

\subsection{Geometric Relationship} 
\label{sec:geometry}

\begin{figure}[H]
  \centering
  \includegraphics[width=\linewidth]{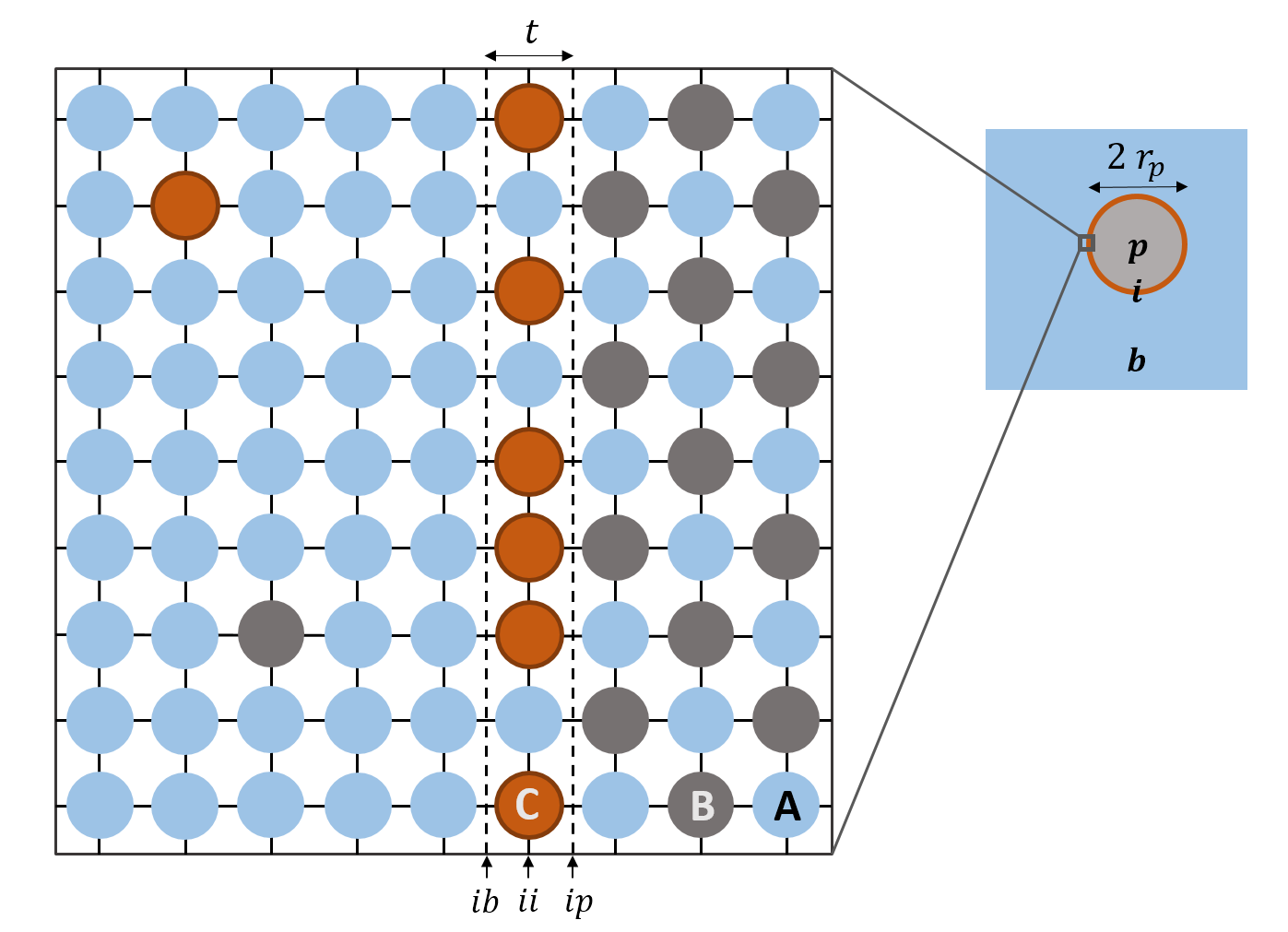}
  \caption{Right: Schematic of a spherical precipitate $p$, of radius
    $r_p$, within the bulk region $b$. The interface region, $i$, is a spherical
    shell of thickness $t$. Left: Region $p$
    comprises an ordered arrangement of atoms $A$ and $B$. Region $b$
    is a random solid-solution of $B$ and $C$ in $A$. The large
    concentration of $C$ in region $i$ illustrates interface solute
    segregation. Bonds in region $i$ correspond to
    bonds between the interface atoms, $ii$, between atoms of $i$ and
    $b$, $ib$, and between atoms of $i$ and $p$, $ip$.}
  \label{fig:geometry}
\end{figure}

The analytical model was developed considering that the interface
atomic region constitutes a single layer of atoms. Thus, the interface
volume fraction, $ f_i $, can be expressed as a fraction of the
precipitate volume fraction, $ f_p $, as $ f_i = \phi f_p $. A
geometric representation of the system configuration can be obtained
for these variables under certain assumptions of the shape and size of
the precipitates. For a spherical morphology with equi-sized
precipitates of radius $r_p$, the interface region is a spherical
shell, of thickness $t$ surrounding the precipitate; this is
illustrated in Fig. \ref{fig:geometry}. Here, $ t $ is taken as $0.25$
nm, which is characteristic of the atomic length scale in
crystals. From geometric dependence of the volumes of the sphere and
the encompassing spherical shell, a relation between the precipitate
size $r_p$ and the fraction of interface with respect to the
precipitate, $\phi$, is obtained as,

\begin{equation} 
\label{eq:rp_main} 
r_p = \frac{3t}{\phi}.
\end{equation}

\subsection{Equilibrium Conditions} \label{sec:eq_cond}

The equilibrium configuration of the ternary system, for a given set
of global compositions $(x_{\circ}, y_{\circ})$, temperature (T) and parametric
values $(\omega_b, \omega_i, \delta_i, z, x_p^{if})$ can be obtained
by minimizing the Gibbs free energy function for the system
(Eq. \ref{eq:dG_tern_main}) with respect to $x_i$, $y_i$, $f_i$ and
$f_p$. The treatment of equilibrium
between the the bulk, the precipitate and the interface regions is
similar to that for equilibrium between three distinct
phases. However, the interface region is different from a phase in
that it cannot exist independent of the bulk and the precipitate
phases. This constraint is incorporated by substituting the relation
$ f_i = \phi f_p $, and by setting minimum and maximum values that
$\phi$ can take. System equilibrium is now obtained by minimizing
$\Delta \bar{G}_{tern}$ with respect to $x_i$, $y_i$, $\phi$ and $f_p$
as:

\begin{equation} \label{eq:equilibrium}
    \frac{\partial \Delta \bar{G}_{tern}}{\partial x_i} = 0, \quad \frac{\partial \Delta \bar{G}_{tern}}{\partial y_i} = 0, \quad \frac{\partial \Delta \bar{G}_{tern}}{\partial \phi} = 0 \quad \text{and} \quad \frac{\partial \Delta \bar{G}_{tern}}{\partial f_p} = 0.
\end{equation}

The equilibrium system configuration obtained in terms of $\phi$ and
$x_b$, $x_i$, $y_b$ and $y_i$ can be represented by the equilibrium
quantities, $r_p^{eq}$ and $\Gamma_i^C$, respectively. (Equilibrium in the binary system is similarly obtained by
minimizing $\Delta \bar{G}_{bin}$ (Eq. \ref{eq:dG_sum_bin}) with
respect to $x_i$, $\phi$ and $f_p$.) Here,
$\Gamma_i^C$ is the excess concentration of $C$ at the interface and
represents the segregation state in the system. $\Gamma_i^C$ for the
ternary system is given by
\cite{finnis1998accessing,sutton1995interfaces}:

\begin{equation} \label{eq:gamma_C}
    \Gamma_i^{C} = \, \frac{1}{N_{avg} \, \Omega^{2/3}} \left[y_i - y_b \ \frac{x_i - \left(1-y_i\right) x_p}{x_b - \left(1 - y_b\right) x_p} \right],
\end{equation}

\noindent where, $N_{avg}$ is a mole of interface atoms, $\Omega$ is
the atomic volume, and $N_{avg} \Omega^{2/3}$ represents the molar
interface area. Defining the interface solute excess per interface
area allows the quantity to be used to compare different
configurational states of the system having different interface volume
fractions.

%% file: ParametricStudy.tex
\section{PARAMETRIC STUDY} 
\label{par_study}

With the thermodynamic model and the conditions for equilibrium
established, we can solve for the equilibrium precipitate size
$(r_p^{eq})$ and interfacial excess $(\Gamma_i^C)$. For the binary
alloy system, the equilibrium precipitate size as a function of global
solute (B) concentration is obtained at either the largest or the
smallest precipitate size allowed by the limits imposed on $\phi$. The
equilibrium states obtained at these limits do not represent a true
equilibrium between the three regions as, in the absence of the
limits, the equilibrium would be between just two regions of the
system. In one case, equilibrium configuration tending towards just
the bulk and precipitate regions arises when the interface energy is
unfavorable compared to other two regions---this is representative of
ideal binary precipitating systems where the precipitates coarsen to
reduce the interfacial energy, provided it is kinetically feasible. In
the other case, equilibrium between bulk and the interface occurs when
the interface energy is favorable; this is obtained when the interface
interaction energy, $\omega_i^{AB}$, is set to a large negative value.

These equilibrium configurations obtained from the binary model can be
rationalized by considering the different regions of the binary system
as phases and invoking the Gibbs phase rule. For a two component
system at constant temperature and pressure, the phase rule states the
the equilibrium between three phases has zero degrees of freedom in
the intensive thermodynamic variables (i.e. chemical potentials of the
components, which in the present model relates to the
concentrations). This means that true equilibrium between the three
regions, represented by a precipitate size within the imposed limits,
can only be obtained at a unique global composition. This is of
limited interest to us, and hence we turn our attention to the ternary
system. For a three component system, as per the phase rule,
equilibrium between three phases is possible with one degree of
freedom in the intensive variables (i.e. concentrations). This
additional degree of freedom is due to the presence of $C$ in the
system. Indeed, we obtain a range of equilibrium precipitate sizes
over a range of global compositions ($x_{\circ}$ or $y_{\circ}$) from the ternary
model; these results are presented below.

The equilibrium system configurations, $r_p^{eq}$ and $\Gamma_i^C$, for
the ternary alloy system are presented in this section as functions of $T$, 
$x_{\circ}$ and $y_{\circ}$, and the interface energy parameters, $\omega_i^{AB}$, 
$\omega_i^{BC}$, $\omega_i^{AC}$, $\delta_i^{AA}$, $\delta_i^{BB}$ and
$\delta_i^{CC}$. Equilibrium values of $x_i$, $y_i$, $f_p$ and $\phi$ are 
obtained by
minimization of $\Delta \bar{G}_{tern}$ (Eq. \ref{eq:dG_tern_main}) using an interior-point optimization routine \cite{matlab2014,fmincon_url}. The 
optimized variables for a given set of parameters are then expressed in terms 
of $\Gamma_i^C$ and $r_p^{eq}$ through relations given in Eqs. \ref{eq:gamma_C} 
and \ref{eq:rp_main}, respectively.

We take Mg-Sn-Zn as an example system for this study as recent work by Liu et al. \cite{liu2016interphase,liu2017zn}
suggests a thermodynamic basis for Zn segregation to the ternary solid-solution/Mg$_2$Sn interface. Accordingly, the values for bulk
interaction parameters defining the Mg-Sn-Zn solid-solution phase are
used, and $\Delta \bar{G}_f^{A_mB_n}$ corresponds to the formation energy of Mg$_2$Sn of $-24.5$ kJ/mol \cite{meng2010thermodynamic} (Supplemental Materials \ref{SI2:therm_data}). In the following study, except for
the particular parameter whose values are varied, all other parameters
are set to default values listed in Table \ref{table:param_def}.
The values of interface interaction parameters,
$\omega_i^{kl}$, are chosen to favor the presence of Zn at the interface,
and the interface penalty parameters, $\delta_i^{kk}$, are chosen
to be positive to represent the energy penalty associated with an
interface. The magnitude of $\delta_i^{AA}$ is based on an estimate of
the average interface energy of Mg/Mg$_2$Sn interface (see Supplemental Materials \ref{SI3:delta}).

\begin{table}[H]
\begin{center}
\caption{Default values are presented for: bulk interaction ($\omega_b$), interface interaction ($\omega_i$) and interface penalty ($\delta_i$) parameters; coordination numbers ($z$); temperature ($T$); global concentration of $B$ ($x_{\circ}$); and precipitate $B$ concentration adjacent to the interface ($x_p^{i/f}$).}
  \begin{tabular}{|c | c | c | c | c | c | c | c | c |  c | c | c | c |  c |  c | c | }
    \hline
    \multicolumn{3}{|c|}{\centering $\omega_b$ (kJ/mol)} & \multicolumn{3}{|c|}{\centering $\omega_i$ (kJ/mol)} & \multicolumn{3}{|c|}{\centering $\delta_i$ (kJ/mol)} & \multicolumn{4}{|c|}{\centering $z$} & \multicolumn{1}{|c|}{$T$} & \multicolumn{1}{|c|}{$x_{\circ}$} & $x_p^{i/f}$ \\
    \hline
    {\centering $\omega_b^{AB}$} & {\centering $\omega_b^{BC}$} & {\centering $\omega_b^{AC}$} & {\centering $\omega_i^{AB}$} & {\centering $\omega_i^{BC}$} & {\centering $\omega_i^{AC}$} & {\centering $\delta_i^{AA}$} & {\centering $\delta_i^{BB}$} & {\centering $\delta_i^{CC}$} & $z_b$ & $z_{ii}$ & $z_{ib}$ & $z_{ip}$ & {\centering (K)} & {\centering (at.\%)} & {\centering (at.\%)} \\
        \hline
    {\centering -2.03} & {\centering +2.54} & {\centering -0.11} & {\centering -2.03} & {\centering +2.54} & {\centering -10} & {\centering 3.5} & {\centering 3.5} & {\centering 3.5} & 12 & 6 & 3 & 3 & {\centering 300}  & 2.2 & 33.33 \\
    \hline
  \end{tabular}
  \label{table:param_def}
\end{center}
\end{table}

\input{param_compositions}

\input{param_omega}

\input{param_delta}

\input{param_temp}

%% file: param_compositions.tex
\subsection{Global solute concentrations, $x_{\circ}$ and $y_{\circ}$}

\begin{figure}[H]
  \centering
  \includegraphics[width=\linewidth]{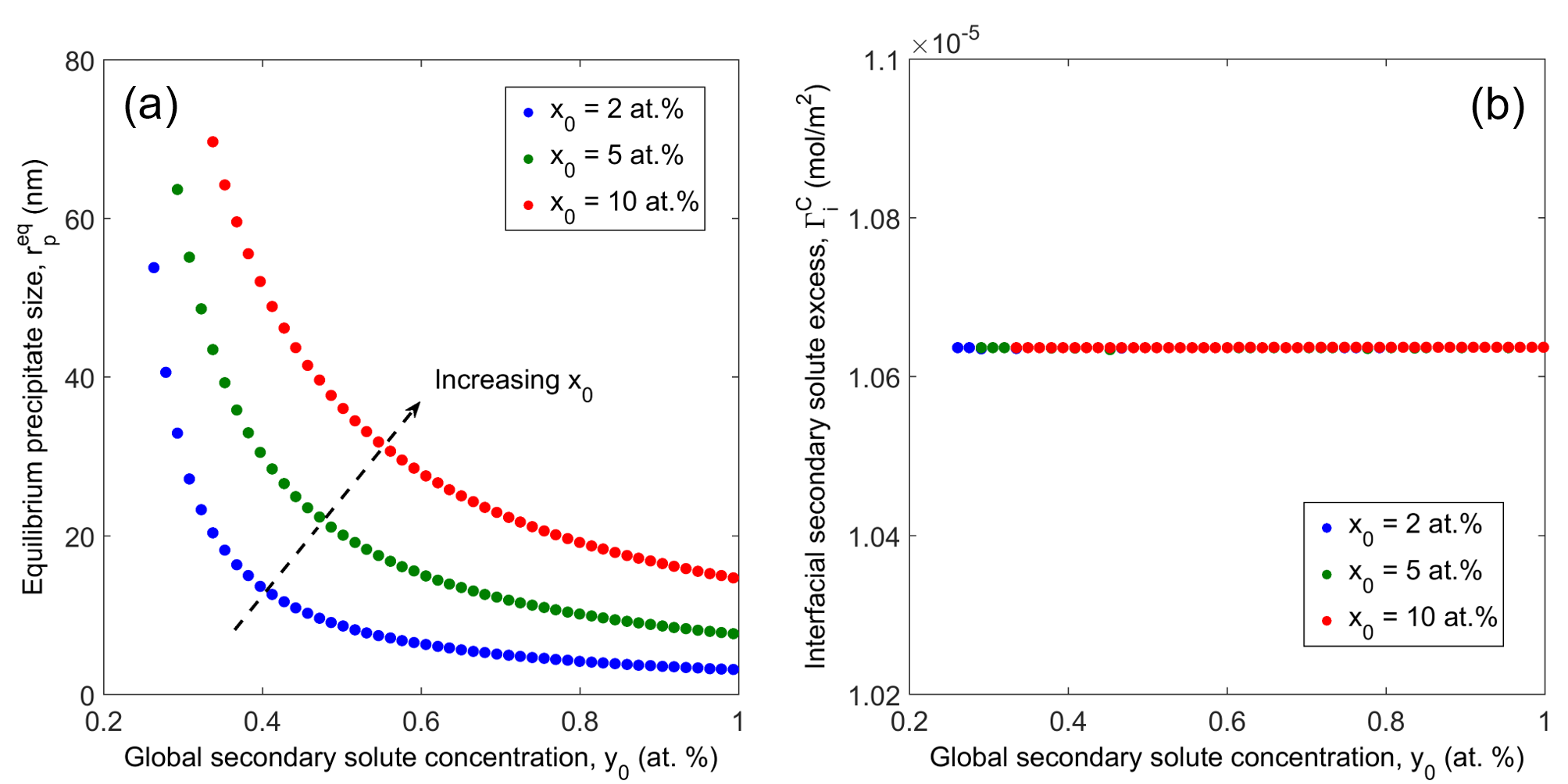}
  \caption{Variation of (a) equilibrium precipitate size and (b) equilibrium
  interfacial secondary solute excess with global solute
  concentrations of B ($x_{\circ}$) and C ($y_{\circ}$). The default values of the
  other relevant parameters are $\omega_b^{AB}=-2.03$,
  $\omega_b^{BC}=+2.54$, $\omega_b^{AC}=-0.11$, $\omega_i^{AB}=-2.03$,
  $\omega_i^{BC}=+2.54$, $\omega_i^{AC}=-10$, $\delta_i^{AA}=3.5$,
  $\delta_i^{BB}=3.5$, $\delta_i^{CC}=3.5$ and $T=300$K (the units of
  $\omega$'s and $\delta$'s is kJ/mol)}.
  \label{fig:x0:req}
\end{figure}

The equilibrium configurations are first analyzed by varying the
concentrations of components B ($x_{\circ}$) and C ($y_{\circ}$) in
the ternary alloy system. Default values listed in Table
\ref{table:param_def} are used for the remaining parameters. In the
range of global concentrations analyzed in this study, the equilibrium
radius, $r_p^{eq}$, of the precipitate was found to be of the order of
a few tens of nanometers. As shown in Fig.\;\ref{fig:x0:req}(a), at
dilute concentrations of solute $C$, a small increase in $y_{\circ}$
results in a dramatic decrease in $r_p^{eq}$. Since the interaction
energy parameters remain constant in this analysis, equilibrium
concentrations in the bulk and the interface regions do not change by
much with a change in the global solute concentrations ($x_{\circ}$ or
$y_{\circ}$). This is reflected by the constant value of $\Gamma_i^C$
in Fig. \ref{fig:x0:req}(b).

$\Gamma_i^C$ for the default parameters corresponds to a large
interfacial concentration of $C$ atoms ($\sim 53$ at.\%) and a low
bulk $C$ concentration ($\sim 0.15$ at.\%), and thus represents a
strong segregation of $C$ atoms to the interface. From an initial
non-equilibrium state of uniform concentration in the bulk and the
interface, segregation of $C$ to the interface reduces the interface
free energy and thus the overall free energy of the system. As
$\omega_i^{AC}$ is assigned a highly negative interaction energy, the
interfacial energy is reduced by maximizing the number of
energetically favorable $AC$ bonds at the interface---this is achieved
at close to equiatomic interface concentration of $A$ and $C$ atoms.

As stated earlier, composition in the bulk
solid-solution, the interface region and the precipitate remain almost
constant as we change the solute concentration $x_{\circ}$ and
$y_{\circ}$. Therefore, the volume fractions of the precipitate and
the interface regions vary to accommodate for the variations in global
solute compositions (this is similar to the lever-rule calculations in
a binary eutectic-alloy). An increase in $y_{\circ}$ while keeping
$x_{\circ}$ fixed will result in an increase in the volume fraction
of interfaces $(f_i)$ at a constant value of precipitate volume
fraction $(f_p)$. This is accomplished by reducing the size and
increasing the number of precipitates, thus increasing the total
interfacial volume fraction. Conversely, increasing $x_{\circ}$ at a
fixed $y_{\circ}$ will result in an increase in $f_p$ while $f_i$
remains constant. This is accomplished by reducing $\phi$ since
$f_i = \phi f_p$, which in-turn results in an increase in $r_p^{eq}$
according to Eq. \ref{eq:rp_main}.



%% file: param_omega.tex
\subsection{Interface interaction energy parameters,
  $\omega_i^{kl}$}

\begin{figure}[H]
  \centering
  \includegraphics[width=0.8\linewidth]{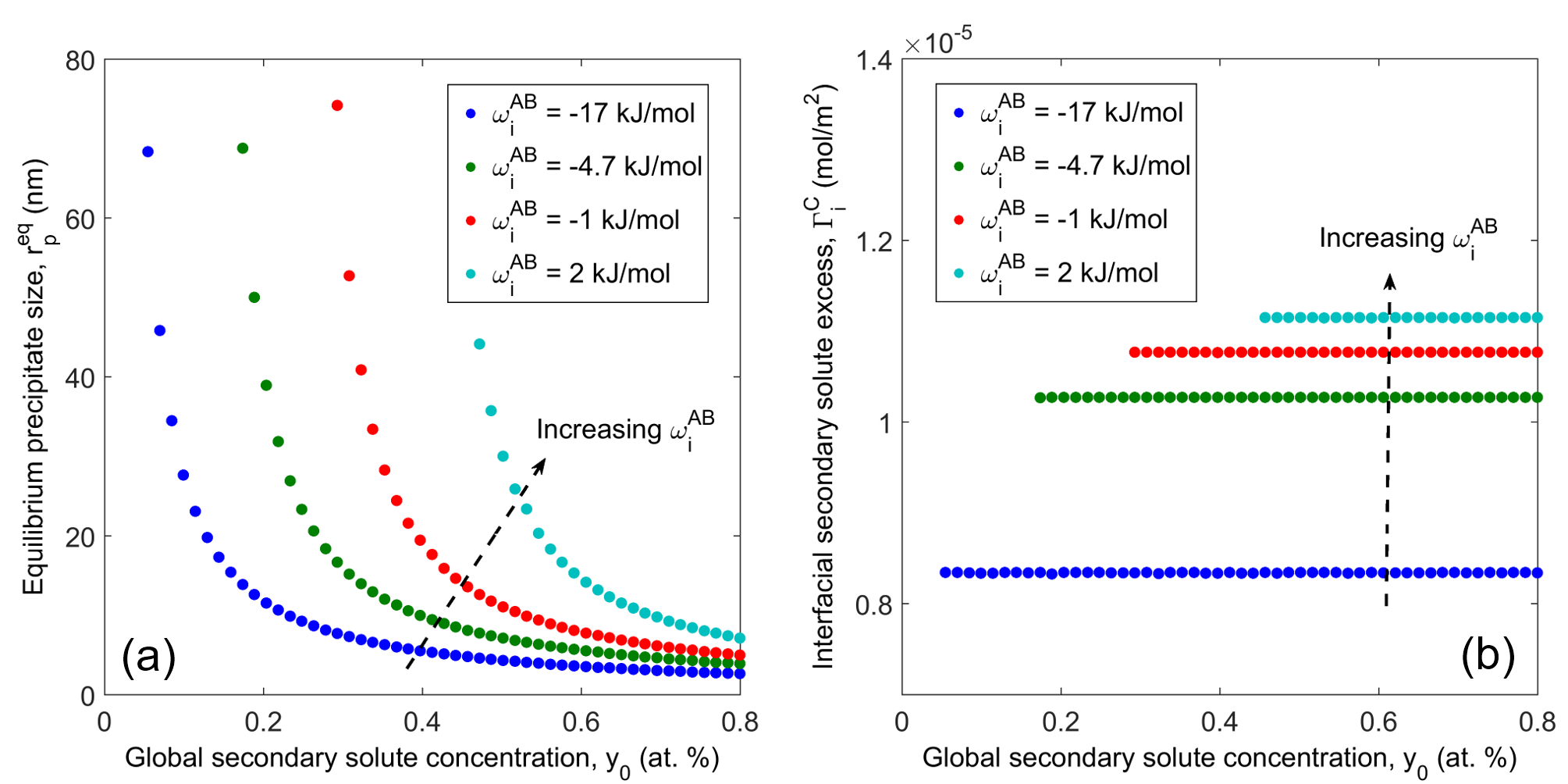}
  \caption{Effect of parametric variation of $\omega_i^{AB}$ on: (a)
    equilibrium precipitate size versus $y_{\circ}$; (b) equilibrium
    interfacial secondary solute excess versus $y_{\circ}$. The default
    values of the other relevant parameters are $\omega_b^{AB}=-2.03$,
    $\omega_b^{BC}=+2.54$, $\omega_b^{AC}=-0.11$,
    $\omega_i^{BC}=+2.54$, $\omega_i^{AC}=-10$, $\delta_i^{AA}=3.5$,
    $\delta_i^{BB}=3.5$, $\delta_i^{CC}=3.5$, $T=300$K,
    $x_{\circ} = 2.2$ at.\% (the units of $\omega$'s and $\delta$'s is
    kJ/mol)}
  \label{fig:wiAB}
\end{figure}

\begin{figure}[H]
  \centering
  \includegraphics[width=0.8\linewidth]{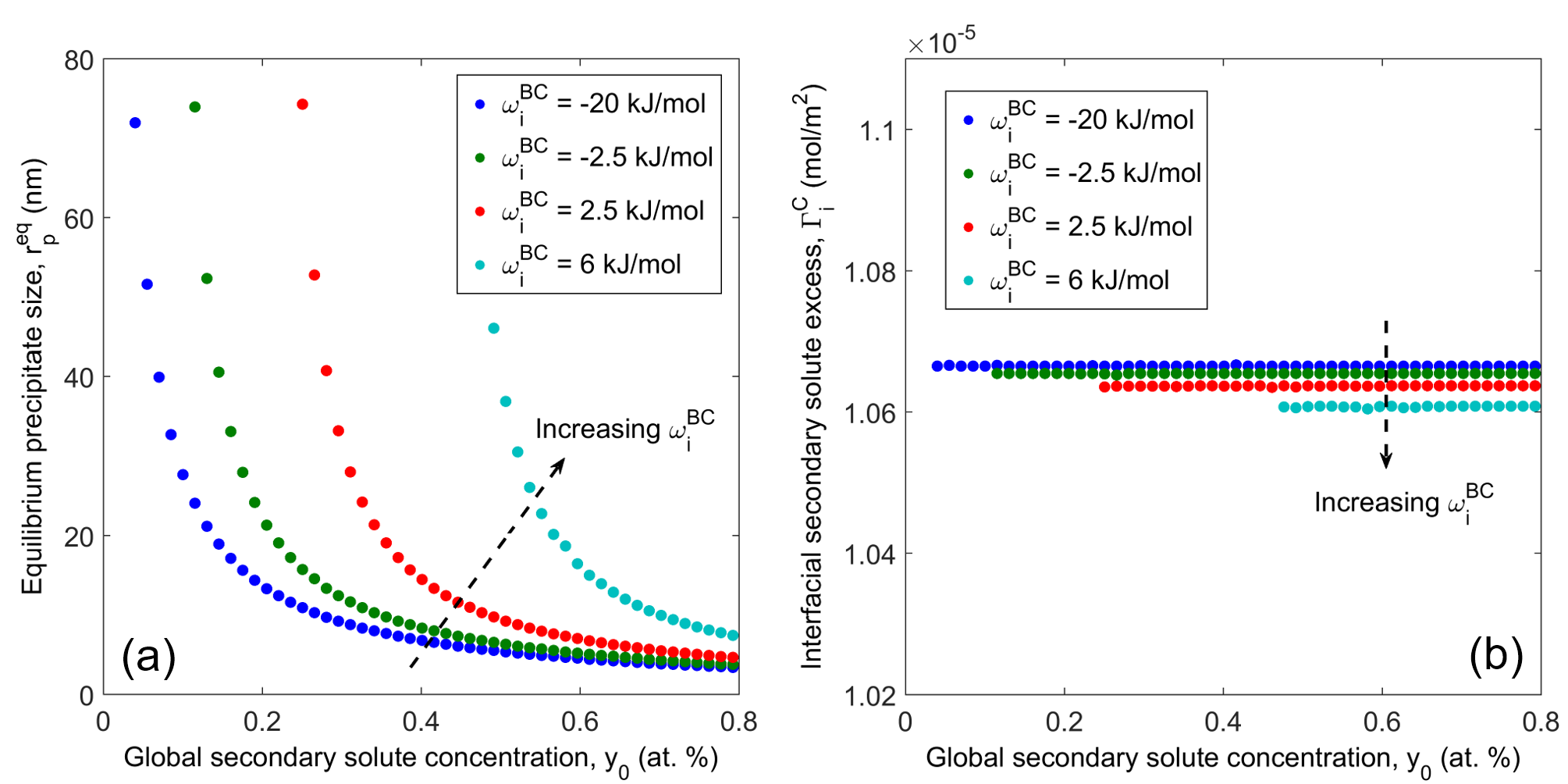}
  \caption{Effect of parametric variation of $\omega_i^{BC}$ on: (a)
    equilibrium precipitate size versus $y_{\circ}$; (b) equilibrium
    interfacial secondary solute excess versus $y_{\circ}$. The default
    values of the other relevant parameters are $\omega_b^{AB}=-2.03$,
    $\omega_b^{BC}=+2.54$, $\omega_b^{AC}=-0.11$,
    $\omega_i^{AB}=-2.03$, $\omega_i^{AC}=-10$, $\delta_i^{AA}=3.5$,
    $\delta_i^{BB}=3.5$, $\delta_i^{CC}=3.5$, $T=300$K,
    $x_{\circ} = 2.2$ at.\% (the units of $\omega$'s and $\delta$'s is
    kJ/mol)}
  \label{fig:wiBC}
\end{figure}

\begin{figure}[H]
  \centering
  \includegraphics[width=0.8\linewidth]{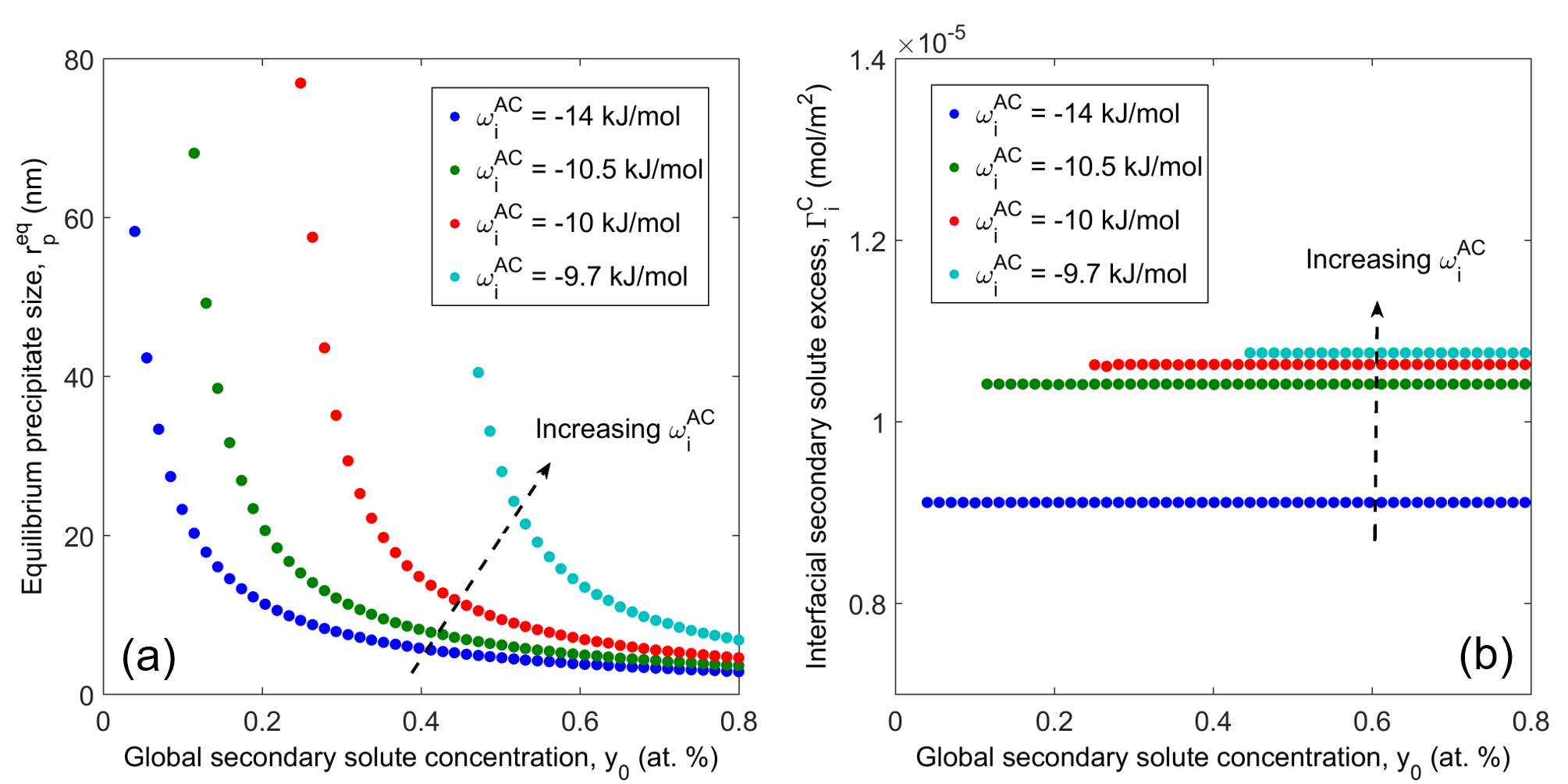}
  \caption{Effect of parametric variation of $\omega_i^{AC}$ on: (a)
    equilibrium precipitate size versus $y_{\circ}$; (b) equilibrium
    interfacial secondary solute excess versus $y_{\circ}$. The default
    values of the other relevant parameters are $\omega_b^{AB}=-2.03$,
    $\omega_b^{BC}=+2.54$, $\omega_b^{AC}=-0.11$,
    $\omega_i^{AB}=-2.03$, $\omega_i^{BC}=+2.54$, $\delta_i^{AA}=3.5$,
    $\delta_i^{BB}=3.5$, $\delta_i^{CC}=3.5$, $T=300$K,
    $x_{\circ} = 2.2$ at.\% (the units of $\omega$'s and $\delta$'s is
    kJ/mol)}
  \label{fig:wiAC}
\end{figure}

To understand the influence of the interface interaction parameters,
the equilibrium precipitate radius $r_p^{eq}$ and the interfacial
solute excess $\Gamma_i^C$ are plotted as a function of the global
secondary solute concentration $y_{\circ}$ as $\omega_i^{AB}$,
$\omega_i^{BC}$ and $\omega_i^{AC}$ are varied in Figs.
\ref{fig:wiAB}, \ref{fig:wiBC} and \ref{fig:wiAC}, respectively. In
general, an unlike bond of type $kl$ is preferred at the interface
over the bulk if
$\omega_i^{kl} + \frac{\delta_i^{kk} + \delta_i^{ll}}{2} <
\omega_b^{kl}$.
Thus, a large negative value of $\omega_i^{kl}$ allows a large volume
fraction of interface to exist at equilibrium by promoting segregation
of solutes that form $kl$ bonds to the interface. Equilibrium
precipitates of nanoscale sizes are therefore achieved for
$\omega_i^{AC} = -10 $ kJ/mol through segregation of $C$ to the
interface and formation of energy reducing $AC$ bonds. As shown in
Figs. \ref{fig:wiAB}(a), \ref{fig:wiBC}(a) and \ref{fig:wiAC}(a), the
equilibrium precipitate size increases with increasing $\omega_i^{kl}$
to account for the increase in the interfacial energy. In the
following, as we justify the observed trends for $\Gamma_i^C$, it is
important to note that almost all of the $B$ atoms are present in the
precipitate due to the large driving force for intermetallic
precipitation. With this in mind, the trends for interfacial solute
excess can be understood as follows:

\begin{itemize}
\item In Fig. \ref{fig:wiAB}b, it is shown that $\Gamma_i^C$ increases
  with increasing $\omega_i^{AB}$. This is because $A$ atoms
  de-segregate out of the interface to reduce the fraction of
  increasingly unfavorable $AB$ bonds in the $ip$ transition
  region. Also, the $C$ atoms substitute interfacial $A$ atoms
  resulting in an increase in the favorable interfacial $AC$ bonds (in
  the transition region), and hence, $\Gamma_i^C$ increases with
  increasing $\omega_i^{AB}$.

\item Similar to the trend in $\omega_i^{AB}$, an increase in
  $\omega_i^{BC}$ will result in a reduction of the number of
  interfacial $BC$ bonds. As most of the $B$ atoms are present in the
  precipitate, the $BC$ bonds are present in the $ip$ transition region.
Therefore, the only way to reduce the number of $BC$ bonds in
  the $ip$ region is by removing the $C$ atoms from the
  interface. This results in a reduction in $\Gamma_i^C$ as shown in
  Fig. \ref{fig:wiBC}(b).

\item An increase in $\omega_i^{AC}$ results in an increase in
  $\Gamma_i^C$ as shown in Fig. \ref{fig:wiAC}(b). While this trend is
  non-intuitive, it is not entirely surprising given the relative
  magnitudes of the $\omega_i$ parameters. The increase in
  $\Gamma_i^C$ with increasing $\omega_i^{AC}$ can be rationalized by
  fixing the volume fraction of the interface $f_i$. This fixes the
  total number of interface bonds (of type $ii$, $ip$ and $ib$) in the
  system. When $\omega_i^{AC}$ is increased, the system requires a
  larger number of interfacial $AC$ type bonds to stabilize the same
  volume fraction $f_i$. This is because the default values for
  $\omega_i^{AB} = -2.03$ kJ/mol and $\omega_i^{BC} = +2.54$ kJ/mol
  are much larger than the values of $\omega_i^{AC}$ explored in this
  study ($-14$, $-10.5$, $-10$ and $-9.7$ kJ/mol). Therefore, in order
  to stabilize the required volume fraction of interfaces in the
  system, $\Gamma_i^C$ has to be increased if $\omega_i^{AC}$ is
  increased.
\end{itemize}

In summary, increasing $\omega_i^{kl}$ displaces $r_p^{eq}$
vs. $y_{\circ}$ curves to larger precipitate sizes; this is presented
in Figs. \ref{fig:wiAB}(a), \ref{fig:wiBC}(a) and \ref{fig:wiAC}(a). As
discussed above, the interface stability corresponding to the new
parametric value of $\omega_i^{kl}$ can be achieved by the system
through a reduction in the unfavorable bond-types and an increase the
favorable bond-types. However, in the absence of the required
concentration of $C$ in the system to partition to the interface, a
larger $\Gamma_i^C$ (for $\omega_i^{AB}$ and $\omega_i^{AC}$
variation) is achieved by decreasing the interface volume fraction;
this is seen by the increase in precipitate size (note that $f_p$
remains constant but the number density of precipitates decreases).



%% file: param_delta.tex
\subsection{Interface energy penalty parameters, $\delta_i^{kk}$}

An energy penalty of $\delta_i^{kk}$ is associated with interface and
transition bonds connected to interface atoms of type $k$. Each
like-bond and unlike-bond has an energy penalty of $ 2 \delta_i^{kk}$
and $\delta_i^{kk} + \delta_i^{ll}$, respectively, and each atom of
type $k$ occupying the interface site has an energy of
$\delta_i^{kk}\frac{z_i}{2}$ in excess of the bulk site (other
energies being equal). To evaluate the effect of the interface energy
penalty parameters on the the equilibrium system configuration,
$\delta_i^{kk}$ for a specific like bond-type is varied, while the
other two like bond-types are fixed at their default value of $3.5$
kJ/mol.

\begin{figure}[H]
  \centering
  \includegraphics[width=0.8\linewidth]{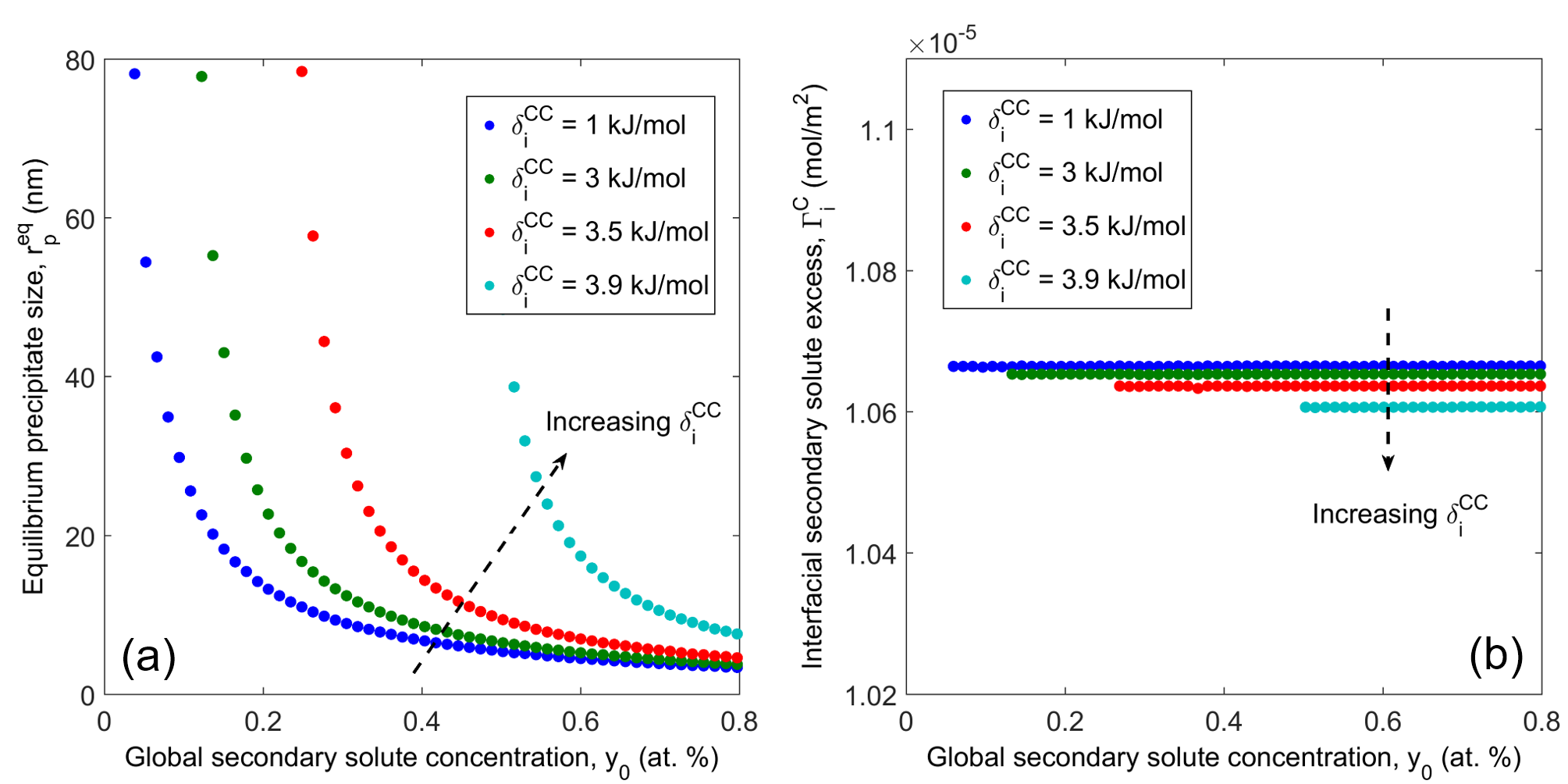}
  \caption{Effect of parametric variation of $\delta_i^{CC}$ on: (a)
    equilibrium precipitate size versus $y_{\circ}$; (b) equilibrium
    interfacial secondary solute excess versus $y_{\circ}$. The default
    values of the other relevant parameters are $\omega_b^{AB}=-2.03$,
    $\omega_b^{BC}=+2.54$, $\omega_b^{AC}=-0.11$,
    $\omega_i^{AB}=-2.03$, $\omega_i^{BC}=+2.54$, $\omega_i^{AC}=-10$,
    $\delta_i^{AA}=3.5$, $\delta_i^{BB}=3.5$, $T=300$K,
    $x_{\circ} = 2.2$ at.\% (the units of $\omega$'s and $\delta$'s is
    kJ/mol).}
  \label{fig:diCC}
\end{figure}

\begin{figure}[H]
  \centering
  \includegraphics[width=0.8\linewidth]{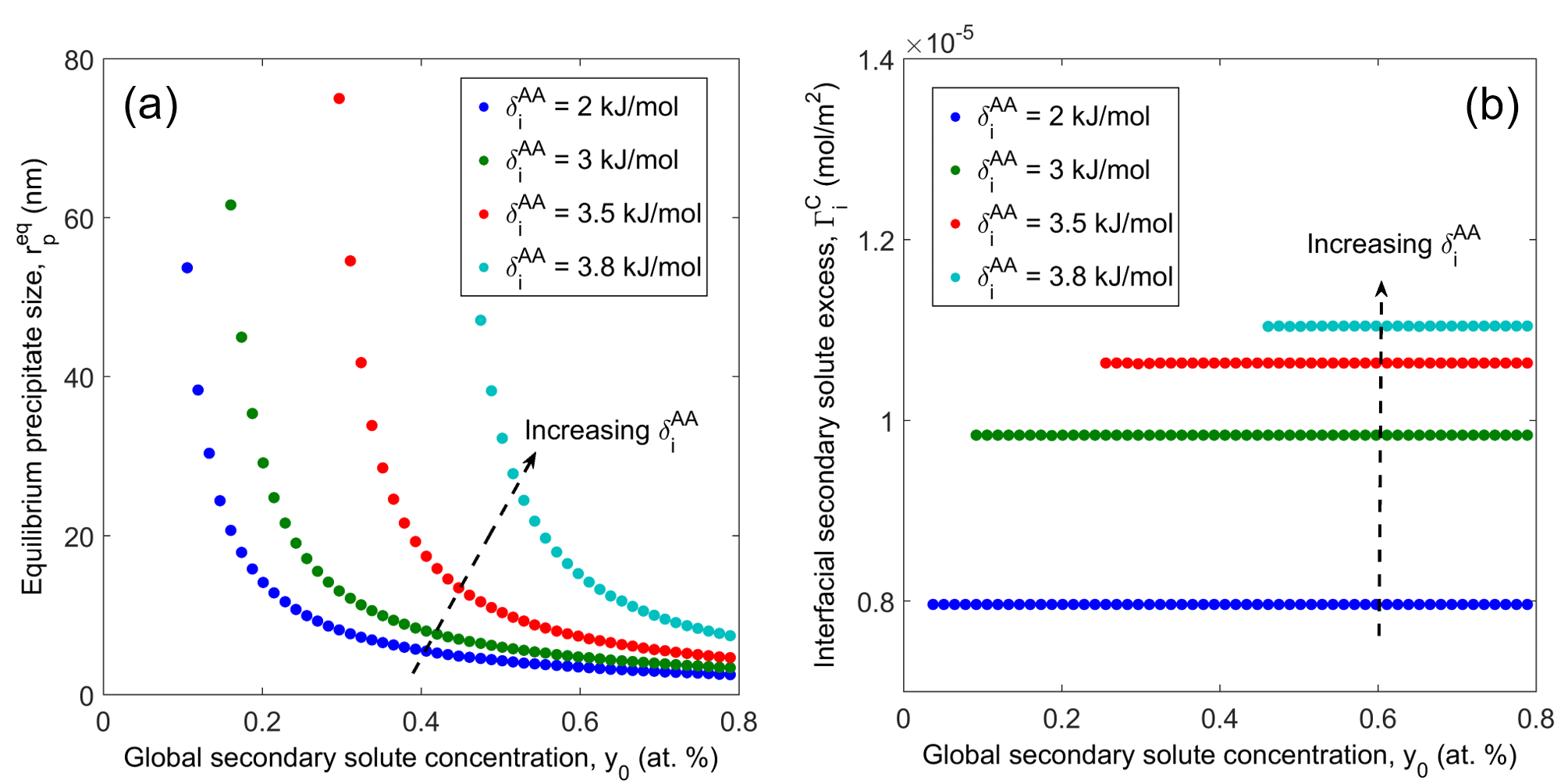}
  \caption{Effect of parametric variation of $\delta_i^{AA}$ on: (a)
    equilibrium precipitate size versus $y_{\circ}$; (b) equilibrium
    interfacial secondary solute excess versus $y_{\circ}$. The default
    values of the other relevant parameters are $\omega_b^{AB}=-2.03$,
    $\omega_b^{BC}=+2.54$, $\omega_b^{AC}=-0.11$,
    $\omega_i^{AB}=-2.03$, $\omega_i^{BC}=+2.54$, $\omega_i^{AC}=-10$,
    $\delta_i^{BB}=3.5$, $\delta_i^{CC}=3.5$, $T=300$K,
    $x_{\circ} = 2.2$ at.\% (the units of $\omega$'s and $\delta$'s is
    kJ/mol).}
  \label{fig:diAA}
\end{figure}

\begin{figure}[H]
  \centering
  \includegraphics[width=0.8\linewidth]{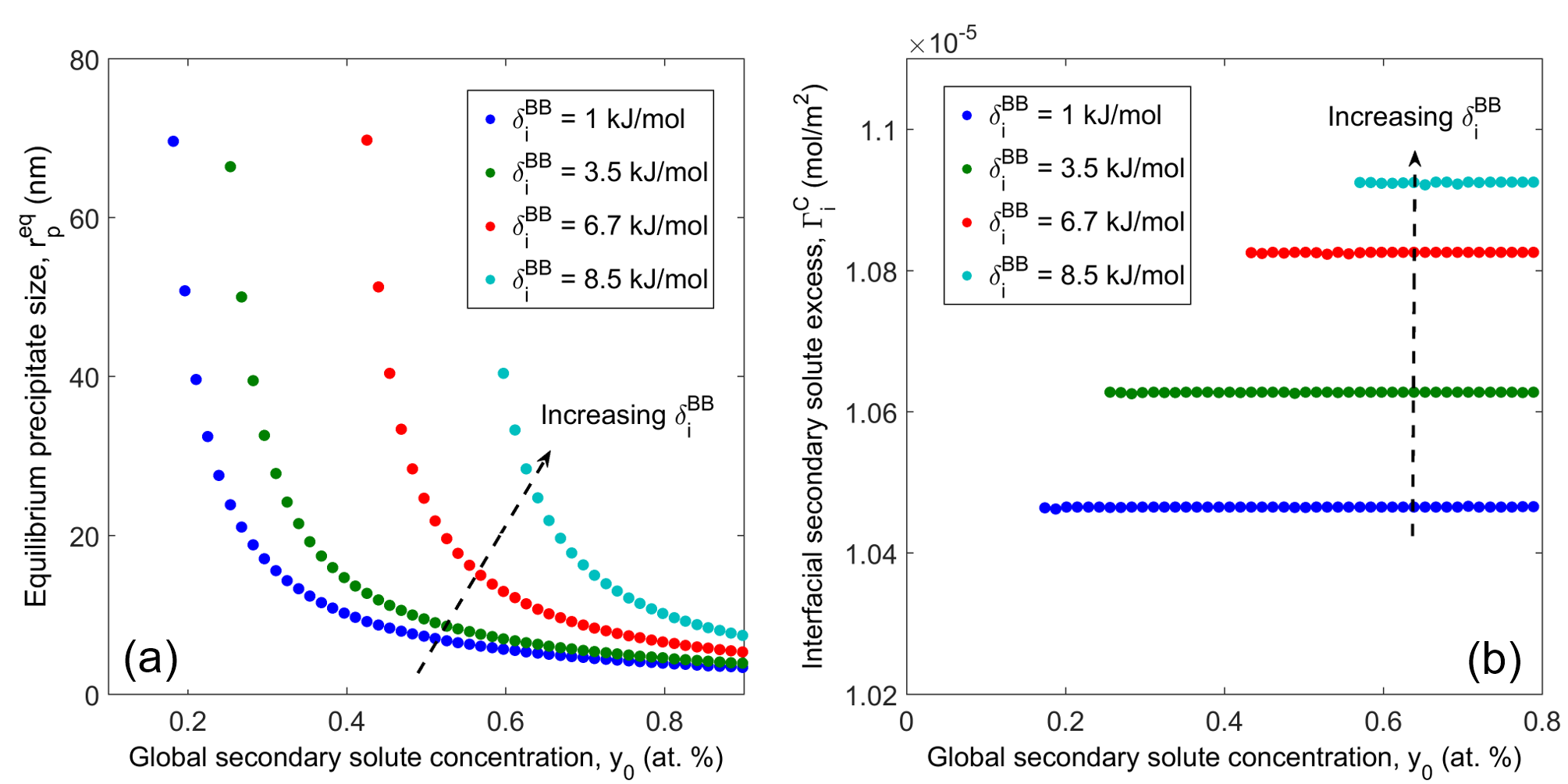}
  \caption{Effect of parametric variation of $\delta_i^{BB}$ on: (a)
    equilibrium precipitate size versus $y_{\circ}$; (b) equilibrium
    interfacial secondary solute excess versus $y_{\circ}$. The default
    values of the other relevant parameters are $\omega_b^{AB}=-2.03$,
    $\omega_b^{BC}=+2.54$, $\omega_b^{AC}=-0.11$,
    $\omega_i^{AB}=-2.03$, $\omega_i^{BC}=+2.54$, $\omega_i^{AC}=-10$,
    $\delta_i^{AA}=3.5$, $\delta_i^{CC}=3.5$, $T=300$K,
    $x_{\circ} = 2.2$ at.\% (the units of $\omega$'s and $\delta$'s is
    kJ/mol).}
  \label{fig:diBB}
\end{figure}

With an increase in $\delta_i^{kk}$, interface stability is maintained
by rejecting from the interface atoms that contribute to excess
interface energy through $kk$-type bonds, and by the segregation to
the interface, atoms that reduce the interface energy through
favorable $kl$ interactions. Thus, an increase in $\delta_i^{CC}$
leads to the de-segregation of $C$ from the interface, as shown by the
decreasing $\Gamma_i^C$ in Fig. \ref{fig:diCC}b, to reduce the overall
number of interface $AC$, $CC$ and $BC$ bonds. Increasing
$\delta_i^{AA}$ leads to rejection of $A$ atoms to reduce the fraction
of $AA$ (mostly) bonds, and segregation of $C$ to increase the
fraction of favorable $AC$ bonds (there is a balance here between the
favorable interaction energy and the energy penalty associated with
$AC$ bonds); this is shown in Fig. \ref{fig:diAA}b. Since the
concentration of $B$ at the interface is negligible, the increase in
interface energy with $\delta_i^{BB}$ corresponds mainly to $AB$ and
$BC$ bonds in the $ip$ transition region. In this case, the interface
energy is reduced by substituting atoms of $A$ with $C$, thereby
increasing $\Gamma_i^C$ (as shown in Fig. \ref{fig:diBB}b) and the
fraction of interface energy reducing $AC$ bonds.

The segregation state of $C$ (i.e. $\Gamma_i^C$) changes with
$\delta_i^{kk}$ to maintain the interface in equilibrium as discussed
above. However, in the absence of sufficient concentration of $C$ in
the system, the equilibrium interface volume fraction decreases as
shown by the shift in $r_p^{eq}$ vs. $y_{\circ}$ curves to larger
precipitate sizes (Figs. \ref{fig:diCC}a, \ref{fig:diBB}a and
\ref{fig:diAA}a). The precipitate can be retained to the same size
(i.e. same $\phi$) by increasing $y_{\circ}$ at a fixed $x_{\circ}$.

%% file: param_temp.tex
\subsection{Temperature}

Dependence of the system's free energy function on temperature arises
from the parameter $T$ coupled to the entropy of mixing terms
corresponding to the bulk and the interface solid-solutions. The
temperature dependence of $\Delta \bar{G}^{ref}_{b,i}$ is ignored and
their values at $300$ K are used as input. The interaction parameters
($\omega$) are temperature-independent in the regular solution
approximation. Thus the variation in the equilibrium system
configuration with $T$ stems primarily from the entropic
contributions.

\begin{figure}[H]
  \centering
  \includegraphics[width=\linewidth]{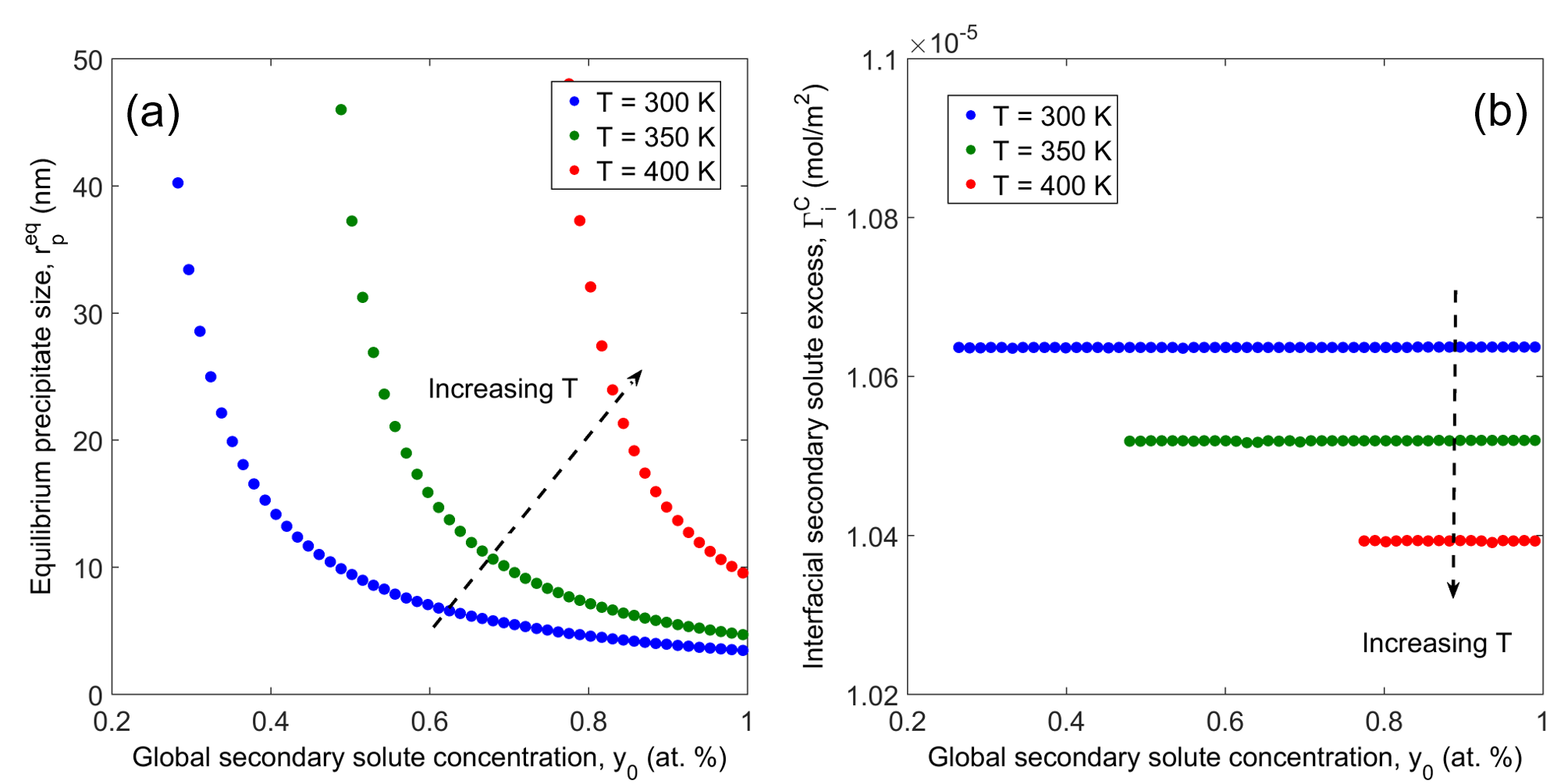}
  \caption{Effect of parametric variation of temperature $(T)$ on: (a)
    equilibrium precipitate size versus $y_{\circ}$; (b) equilibrium
    interfacial secondary solute excess versus $y_{\circ}$. The default
    values of the other relevant parameters are $\omega_b^{AB}=-2.03$,
    $\omega_b^{BC}=+2.54$, $\omega_b^{AC}=-0.11$,
    $\omega_i^{AB}=-2.03$, $\omega_i^{BC}=+2.54$, $\omega_i^{AC}=-10$,
    $\delta_i^{AA}=3.5$, $\delta_i^{BB}=3.5$, $\delta_i^{CC}=3.5$,
    $x_{\circ} = 2.2$ at.\% (the units of $\omega$'s and $\delta$'s is
    kJ/mol).}
  \label{fig:varyT}
\end{figure}

With an increase in temperature, precipitates are stabilized at larger
sizes as shown in Fig. \ref{fig:varyT}(a). Correspondingly,
concentration of $C$ at the interface decreases as shown by the
decrease in the interfacial secondary solute excess in
Fig. \ref{fig:varyT}b. The results correspond to de-segregation of $C$
from the interface with temperature, leading to a reduction in the
equilibrium interface volume fraction, and therefore, an increase in
stable precipitate size. The cause of desegregation is the reduction
in system free energy through an increase in the entropy of the bulk
solid-solution region. While the total entropy of the system is a
combination of configurational entropy of mixing of bulk and interface
solid-solutions, since the volume fraction of the bulk is
significantly greater than that of the interface ($f_b>0.9$), the
system entropy is dominated by the contribution from the bulk (second
term in Eq. \ref{eq:dGb_bin}). Therefore, with increasing temperature,
the free energy of the system is reduced by increasing the
configurational entropy of the bulk, which is accomplished by the
desegregation of $C$ from the interface to the bulk. Free energy
minimization, however, is a trade-off between enthalpic and entropic
contributions. A system with stronger interactions of $C$ atoms at the
interface, over the bulk, will have a lower tendency to de-segregate
if the enthalpic driving force for segregation to the interface is
stronger than the entropic driving force for de-segregation.

%% file: Conclusions.tex
\section{CONCLUSIONS} 
\label{Conclusions}

We developed an analytical model that captures the thermodynamic
stabilization of nano-sized precipitates, through chemically driven
solute segregation to the heterophase interface, in alloy systems
precipitating an intermetallic compound from the solid-solution. In
binary precipitate systems like Mg/Mg$_2$Sn, or in the absence of
secondary solutes with energy reducing interface interactions, the
interface always presents a positive energy penalty which can only be
reduced through precipitate coarsening. While aging treatments can be
optimized to produce an alloy with small-size and high number density
of precipitates to provide beneficial mechanical properties at room
temperature, these precipitates are only kinetically stable. At higher
temperatures, the precipitates coarsen to large sizes (typically to
micron length scales) and drastically degrade structural
performance. On the other hand, in ternary (or quaternary) alloy
systems, where secondary solutes are chosen to have strong
interactions at the heterophase interface, thermodynamic stabilization
through solute segregation can be effective at high temperatures. For
example, results of the ternary model evaluated for Mg-Sn-Zn system
show that precipitates can be stabilized through Zn segregation driven
by thermodynamic interfacial energy minimization.

While this study presents a thermodynamic basis for the experimental
findings \cite{liu2016interphase} of Zn segregation to Mg/Mg$_2$Sn
interface and an associated refinement of Mg$_2$Sn precipitate sizes,
actual values for the interface parameters are unknown. The parametric
study presented in this paper shows a reduction in the equilibrium
precipitate size when the global secondary solute concentration,
$y_{\circ}$, is increased or any of the interface interaction
parameters are decreased. On the other hand, the equilibrium
precipitate size increases with temperature due to the entropic
contributions favoring a more random partitioning of concentration
between bulk and interface regions. While the present model
incorporates chemical interactions, extension to the model is
necessary to capture segregation and precipitate stabilization
resulting from elastic strain energy effects. Together with the input
of interface parameters from methods like density functional theory
calculations, and the inclusion of participate morphology as a
variable, the model will allow a more realistic prediction of
thermodynamically stable precipitate sizes and shapes in ternary alloy
systems.


%% file: SupportingInfo.tex
\begin{widetext}
\begin{center}
  \textbf{\large Supplemental Materials \\ Thermodynamic Stabilization of Precipitates through Interface Segregation: Chemical Effects}
\end{center}

\setcounter{equation}{0}
\setcounter{figure}{0}
\setcounter{table}{0}
\setcounter{page}{1}
\setcounter{section}{0}

\makeatletter

\renewcommand{\theequation}{S\arabic{equation}}
\renewcommand{\thesection}{S\arabic{section}}
\renewcommand{\thefigure}{S\arabic{figure}}
\renewcommand{\bibnumfmt}[1]{[S#1]}
\renewcommand{\citenumfont}[1]{S#1}

\begin{bibunit}[apsrev4-1]


\section{ANALYTICAL TREATMENT}
\label{SI1:Atreat}

\subsection{Binary Model} \label{SI:Atreat:bin}
\subsubsection{Internal Energy Function} \label{SI:Atreat:bin:U}

We consider a binary system consisting of three distinct regions---bulk solid-solution, intermetallic precipitate, and a single-atomic-layer of interface solid-solution between the bulk and the precipitate; these regions of atom occupancy are denoted by $b$, $p$ and $i$,
respectively. The total internal energy of the system in this
configuration, $ U_{sys} $, is the sum of the individual internal
energies of each of the three regions; $U_{sys}$ is defined for the
thermodynamic state at temperature $T$, total volume of the system
$V$, and total number of atoms in the system $N_{\circ}$. Considering that
an atom of a particular type $k$ (i.e. $A$ or $B$), within an atomic region
$reg$, has an average potential energy per atom
$ E_{reg}^k $, the internal energy of the system with $N_{reg}^k$ number of
$k$-type atoms in each region $reg$ is,

\begin{equation}
  \label{eq:int_eng}
  U_{sys} = \sum\limits_{reg = b,i,p} \left(N_{reg}^A E_{reg}^A + N_{reg}^B E_{reg}^B \right).
\end{equation}

The initial or standard configuration of the system, at the same $T$ as the system configuration, is taken as the
pure components in their standard element crystal structures (the
stable form at 298.15 K and $10^5$ Pa). Per-atom energies in this standard state, $std$, is denoted by
$ E_{std}^k$. The internal energy of this unmixed,
interface- and precipitate-free configuration is given by,

\begin{equation}
  \label{eq:int_std}
  U_{std} = N_{total}^A E_{std}^A + N_{total}^B E_{std}^B,
\end{equation}

\noindent where, $ N_{total}^k $ is the total number of atoms of type
$k$ in the system, and is related to $ N_{reg}^k $ by,

\begin{align} \label{eq:atom_cons}
  N_{total}^A =& N_b^A + N_i^A + N_p^A, \nonumber\\
  N_{total}^B =& N_b^B + N_i^B + N_p^B.
\end{align}

The internal energy change, $ \Delta U_{bin} $, for the formation of the
binary system with the three regions $b$, $i$, and $p$ from the
initial configuration of pure components is given by,

\begin{align}
  \label{eq:del_int}
  \Delta U_{bin} = &U_{sys} - U_{std} \nonumber\\
  = &\sum\limits_{reg = b,i,p} \left[N_{reg}^A \left(E_{reg}^A - E_{std}^A \right) + N_{reg}^B \left(E_{reg}^B - E_{std}^B \right) \right].
\end{align}

According to Eq.\ (\ref{eq:del_int}), the reference energies for all
the three regions of the system are $E_{std}^A$ and $E_{std}^B$.
The crystal structures of the solid-solution phases, however, may be
different from that of the standard states, and the thermodynamic parameters of mixing available in thermodynamic literature/database correspond to a reference state that has the same crystal structure as the solid-solution. Therefore, we reformulate terms corresponding to $b$ and $i$ in Eq. \ref{eq:del_int} to obtain
reference states as the pure components with energies and crystal structures characteristic of $b$ and $i$.
The
energies of atoms of type $k$, specific to region $reg$, in their reference state
is denoted by $E_{ref(reg)}^k$. Next, the energy
difference between the reference state of $i$ and the pure component
standard state, $E_{ref(i)} - E_{std}$, is rewritten in terms of $E_{ref(i)} - E_{ref(b)}$ and $E_{ref(b)} - E_{std}$. The reference state for region $p$ is left unchanged as the standard state since the data for formation energies of intermetallics in literature/databases are defined with respect to the standard states. Thus, $\Delta U_{bin}$ is obtained as,

\begin{subequations} \label{eq:dU_def}
  \begin{align} 
    \Delta U_{bin} =& \Delta U^{\,mix}_{b} + \Delta U_{i} + \Delta U^{\,ref}_{b,i} + \Delta U_p, \tag{\ref{eq:dU_def}}\\[20pt]
    \text{where,} \nonumber\\[20pt]
    \Delta U^{\,mix}_{b} =& \left[\left(N_b^A E_b^A + N_b^B E_b^B \right) - \left(N_b^A E_{ref(b)}^A + N_b^B E_{ref(b)}^B \right)\right], \label{eq:dU_defa}\\[20pt]
    \Delta U_{i} =& \left[\left(N_i^A E_i^A + N_i^B E_i^B \right) - \left(N_i^A E_{ref(i)}^A + N_i^B E_{ref(i)}^B \right)\right] \nonumber \\
    &+ \left[N_i^A \left(E_{ref(i)}^A - E_{ref(b)}^A \right) + N_i^B \left(E_{ref(i)}^B - E_{ref(b)}^B \right) \right], \label{eq:dU_def_b}\\[20pt]
    \Delta U^{\,ref}_{b,i} =& \left[N_b^A \left(E_{ref(b)}^A - E_{std}^A \right) + N_b^B \left(E_{ref(b)}^B - E_{std}^B \right) \right] \nonumber \\
    &+ \left[N_i^A \left(E_{ref(b)}^A - E_{std}^A \right) + N_i^B \left(E_{ref(b)}^B - E_{std}^B \right) \right], \label{eq:dU_def_c}\\[20pt]
    \Delta U_p =& \left[\left(N_p^A E_p^A + N_p^B E_p^B \right) - N_p^A E_{std}^A - N_p^B E_{std}^B \right]. \label{eq:dU_def_d}
  \end{align}
\end{subequations}

$\Delta U^{\,mix}_{b}$ and $\Delta U_{i}$ represent the internal energy
change associated with the formation of
solid-solution regions $b$ and $i$, respectively, from the pure component reference state $ref(b)$. By considering
random mixing and pair-wise interactions, the mixing in $b$ and $i$
can be treated similar to the regular
solution models \cite{dehoff2006thermodynamics, trelewicz2009grain}. Thus, Eqs. (\ref{eq:dU_defa}) and (\ref{eq:dU_def_b}) are
redefined from a formalism involving an average per-atom energy
description to a bond energy description based on the
nearest-neighbor pairwise interaction. $N_{b/i}^k$ is expressed in terms of the number of $kl$ bonds, $ N_r^{kl} $,
and $E_{b/i}^k$ in terms of the corresponding bond energies, $ E_{b/i}^{kl} $. The bonding regions ($r$) are designated as follows: $ib$ for bonds between atoms of $i$ and $b$; $ip$ for bonds between atoms of $i$ and $p$; $ii$ for bonds within atomic region $i$; $b$ for bonds within atomic region $b$.

\begin{align} \label{eq:Umix_b}
	\Delta U^{\,mix}_{b} &= \Big(N_b^{AA} E_b^{AA} + N_b^{BB} E_b^{BB} + N_b^{AB} E_b^{AB} \Big) - \left(N_{ref(b)}^{AA} E_{ref(b)}^{AA} + N_{ref(b)}^{BB} E_{ref(b)}^{BB} \right)
\end{align}

\begin{align} \label{eq:Umix_i}
	\Delta U_{i} &= \Big(N_i^{AA} E_i^{AA} + N_i^{BB} E_i^{BB} + N_i^{AB} E_i^{AB} \Big) - \left(N_{ref(i)}^{AA} E_{ref(i)}^{AA} + N_{ref(i)}^{BB} E_{ref(i)}^{BB} \right) \nonumber \\
    &\quad+ \left[N_i^A \left(E_{ref(i)}^A - E_{ref(b)}^A \right) + N_i^B \left(E_{ref(i)}^B - E_{ref(b)}^B \right) \right]
\end{align}

Relation between the number of atoms of type $k$, the number
of bonds of type $kk$ between like atoms and type $kl$ between unlike
atoms, and the number of bonds in the reference state, for atomic regions $b$ and $i$ is given by,

\begin{subequations} \label{eq:atm_rel1}
  \begin{align} 
    N_b^k z_b & = 2N_{ref(b)}^{kk} = 2N_b^{kk} + N_b^{kl}, \label{eq:atm_rel1a}\\
    N_i^k z_i & = 2N_{ref(i)}^{kk} = 2N_i^{kk} + N_i^{kl} \label{eq:atm_rel1b} \nonumber\\
              & = \left(2N_{ii}^{kk} + N_{ii}^{kl} \right) + \left(2N_{ib}^{kk} + N_{ib}^{kl} \right) + \left(2N_{ip}^{kk} + N_{ip}^{kl} \right).
  \end{align} \nonumber
\end{subequations}
\\
Now, expressions for $N_{ref(b)}^{kk}$ from Eq. (\ref{eq:atm_rel1a}) and for $N_{ref(i)}^{kk}$, $N_i^A$ and $N_i^B$ from Eq. (\ref{eq:atm_rel1b}) are substituted in Eq. (\ref{eq:Umix_i}). Also substituting
$E_{ref(r)}^{kk} = E_{r}^{kk}$ gives:

\begin{align} \label{eq:atm_bmix}
    \Delta U^{\,mix}_{b} &= N_b^{AB}\left(E_b^{AB}-\frac{E_b^{AA}+E_b^{BB}}{2}\right)
\end{align}

\begin{align} \label{eq:atm_imix}
    \Delta U_{i} &= \sum\limits_{r = ii,ib,ip} \left[N_r^{AB}\left(E_r^{AB}-\frac{E_r^{AA}+E_r^{BB}}{2}\right) \right. \nonumber \\
    &+ \left.\left(N_r^{AA}+\frac{N_r^{AB}}{2}\right)\frac{2}{z_i}\left(E_{ref(r)}^A-E_{ref(b)}^A\right) + \left(N_r^{BB}+\frac{N_r^{AB}}{2}\right)\frac{2}{z_i}\left(E_{ref(r)}^B-E_{ref(b)}^B\right)\right] \nonumber\\
\end{align}

\noindent In Eq. (\ref{eq:atm_imix}), the energies of transition bonds are considered to be characteristic of interface bonds, i.e. for $r$ = $ii$, $ib$ and $ip$,
$E_{r}^{kk/ kl} = E_{i}^{kk/ kl}$ and
$E_{ref(r)}^{k} = E_{ref(i)}^{k}$. This also applies to the
energy parameters defined below.

The regular solution interaction parameters, capturing the energetics
of formation of unlike bond types from like bond types, characteristic
to $b$ and $i$ (i.e. $ii$, $ib$ and $ip$) are defined as,

\begin{subequations} \label{eq:omega_def}
\begin{align} \label{eq:om_ab} 
	\omega_b^{AB} &= E_b^{AB} - \frac{E_b^{AA}+E_b^{BB}}{2}, \\[10pt]
    \omega_i^{AB} &= E_i^{AB} - \frac{E_i^{AA}+E_i^{BB}}{2}.
\end{align}
\end{subequations}
\\
We define an energy penalty parameter, $\delta_r^{kk}$, that represents the excess energy of interface-type bonds over bulk-type bonds. This is expressed as the difference between per-atom energies of interface and bulk
reference states as,

\begin{align} \label{eq:deli_aa} \delta_i^{AA} =
  \frac{2}{z_i}\left(E_{ref(i)}^A - E_{ref(b)}^A\right), \qquad
  \delta_i^{BB} = \frac{2}{z_i}\left(E_{ref(i)}^B -
    E_{ref(b)}^B\right).
\end{align}
\noindent Alternately,
\begin{align} \label{eq:deli_alt} \delta_i^{AA} =
  E_{i}^{AA} - \frac{z_b}{z_i} E_{b}^{AA}, \qquad
  \delta_i^{BB} = E_{i}^{BB} -
    \frac{z_b}{z_i} E_{b}^{BB}.
\end{align}

\noindent Eq. (\ref{eq:deli_aa}) for any bond-region $r$ is represented by $\delta_r^{kk}$ ($\delta_b^{kk} = 0$) and by the coordination number $z_r$. The
energy difference between the pure component reference state of $b$
and the pure component standard state is defined as,

\begin{equation} \label{eq:delu_std} \Delta U_{ref(b)}^A = E_{ref(b)}^A - E_{std}^A, \qquad \Delta U_{ref(b)}^B = E_{ref(b)}^B - E_{std}^B.
\end{equation}
\\
Using the definitions of Eqs. (\ref{eq:omega_def}), (\ref{eq:deli_aa}) and
(\ref{eq:delu_std}), Eqs. (\ref{eq:atm_bmix}) and (\ref{eq:atm_imix}) can be rewritten as:

\begin{align} \label{eq:Ubmix-bonds}
  \begin{split}
    \Delta U^{\,mix}_{b} = N_b^{AB}\omega_b^{AB}
  \end{split}
\end{align}

\begin{align} \label{eq:Uimix-bonds}
  \begin{split}
    \Delta U_{i} = \sum\limits_{r = ii,ib,ip} & \left[ N_r^{AB}\omega_r^{AB} + \left(N_r^{AA}+\frac{N_r^{AB}}{2}\right) \delta_r^{AA} + \left(N_r^{BB}+\frac{N_r^{AB}}{2}\right)\delta_r^{BB} \right]
  \end{split}
\end{align}

Supposing uniform number density of atoms within the system and equal
atomic volumes of the components, the global or system concentration
of $B$ ($x_{\circ}$) can be expressed in terms of region-specific
concentrations of $B$ ($x_b$, $x_i$ and $x_p$) and region-specific
volume fractions ($f_i$ and $f_p$) through Eq. (\ref{eq:cons_x}). Since the precipitate region is chosen to be stoichiometric,
$x_p = n/m$ for an $A_mB_n$ intermetallic.

\begin{equation} \label{eq:cons_x}
  x_{\circ} = x_b \left(1-f_i-f_p \right) + x_i f_i + x_p f_p \\
\end{equation}

The number of atoms of a given
type and region affiliation, $N_{reg}^k$, in
Eq. (\ref{eq:dU_def_c}) can be expressed as,

\begin{align} \label{eq:atom_cons1}
  \begin{aligned}
    N_b^A &= \left(1-f_i-f_p\right) N_{\circ} \left(1-x_b\right), \\ N_b^B
    &= \left(1-f_i-f_p\right) N_{\circ} x_b,
  \end{aligned}
    &&
       \begin{aligned}
         N_i^A &= f_i N_{\circ} \left(1-x_i\right), \\ N_i^B &= f_i N_{\circ} x_i.
       \end{aligned}
\end{align}

Expressions for $N_r^{kl}$ in Eqs. (\ref{eq:Ubmix-bonds}) and (\ref{eq:Uimix-bonds}) can be obtained
using statistical consideration of the existential bond probability of
$kl$ (like and unlike) bonds among the total number of all bond types in the
given region $(N_r^{bonds})$ as,

\begin{equation} \label{eq:bond_cons1} N_r^{kl} = N_r^{bonds}P_r^{kl}.
\end{equation}

\noindent Relations for $N_r^{bonds}$ and $P_r^{kl}$ are listed in Table \ref{table:binary_main} of the main paper. The bond probabilities are derived for $b$ and $i$ regions based on
random site occupancy of atoms---the probability of occupancy of a
lattice site by a component $k$ is the concentration of $k$. 

The interface atoms are considered to contribute a part of
their bond co-ordination, $z_{ib}$, to $ib$ transition
region, and $z_{ip}$ to $ip$ transition
region. The rest of the interface bond co-ordination, $z_{ii}$,
connects interface atoms lying within the interface atomic region. The
$ip$ transition bonds connect interface atoms with
the precipitate atoms located at the layer of the
precipitate region that is adjacent to the transition region. The
concentration at this precipitate layer is uniquely defined by $x_p^{i/f}$.

The summation in Eq. (\ref{eq:Uimix-bonds}) is expanded over $r$, and the
number of terms in the expansion is reduced by substituting
$ \delta_b^{kk} = 0 $. Since $ib$ and $ip$ transitional bonds are assigned bond
energies characteristic of the interface ($i$),

\begin{equation}\label{eq:intp_cons1}
  \omega_i^{kl} = \omega_{ib}^{kl} = \omega_{ip}^{kl} = \omega_{ii}^{kl}, \qquad \delta_i^{kk} = \delta_{ib}^{kk} = \delta_{ip}^{kk} = \delta_{ii}^{kk}.
\end{equation}

Eq. (\ref{eq:dU_def_d}) represents the internal
energy for the formation of $A_mB_n$ from the standard state. The first part of this term is the internal energy of $A_mB_n$ and is defined by $U_f^{A_mB_n}$
as,

\begin{equation} \label{eq:int_formula1} \frac{U_f^{A_m B_n}}{m+n} =
  x_p^A E_p^A + x_p^B E_p^B
\end{equation}

The energy parameters of Eqs. (\ref{eq:intp_cons1}) and (\ref{eq:int_formula1}) and the relations from Table \ref{table:binary_main} are substituted into $\Delta U^{\,mix}_{b}$ (Eq. \ref{eq:Ubmix-bonds}), $\Delta U_{i}$ (Eq. \ref{eq:Uimix-bonds}), $\Delta U^{\,ref}_{b,i}$ (Eq. \ref{eq:dU_def_c}) and $\Delta U_p$ (Eq. \ref{eq:dU_def_d}). Rearranging the resulting expression, considering $N_{\circ}$ as a mole of atoms in the system, and redefining the energy parameters per mole yields the final expression for the internal energy function of binary system, $\Delta \bar{U}_{bin}$, per mole of atoms (represented by the bar over $U$) as:

\begin{subequations} \label{eq:dU_binary}
  \begin{align} 
    \Delta \bar{U}_{bin} =& \, \Delta \bar{U}^{\,mix}_{b} + \Delta \bar{U}_{i} + \Delta \bar{U}^{\,ref}_{b,i} + \Delta \bar{U}_p, \tag{\ref{eq:dU_binary}}\\[20pt]
    \Delta \bar{U}_i =& \, \Delta \bar{U}^{\,mix}_{i} + \Delta \bar{U}_{ib} + \Delta \bar{U}_{ip}, \\[20pt]
    \text{where,} \nonumber\\[20pt]
    \Delta \bar{U}^{\,mix}_{b} =& \left[\omega_b^{AB} \left(1 - x_b \right) x_b \right] \left(1 - f_{i} - f_p \right) z_b, \label{eq:dU_binary_a}\\[20pt]
    \Delta \bar{U}^{\,mix}_{i} =& \left[\omega_{i}^{AB} \left(1 - x_{i} \right) x_{i} \right] f_{i} z_{ii} + \left[\delta_i^{AA}\left(1-x_{i}\right) + \delta_i^{BB}x_{i} \right]f_{i} \frac{z_{ii}}{2}, \label{eq:dU_binary_b}\\[20pt]
    \Delta \bar{U}_{ib} =& \, \omega_{i}^{AB} \left[ \left(1-x_b \right)x_{i} + \left(1-x_{i} \right)x_b \right] f_{i} \frac{z_{ib}}{2} \nonumber\\
    &+ \left[\delta_i^{AA}\left(1-x_{i} + 1-x_{b} \right) + \delta_i^{BB}\left(x_{i}+x_b \right) \right]f_{i} \frac{z_{ib}}{4}, \label{eq:dU_binary_c}\\[20pt]
    \Delta \bar{U}_{ip} =& \, \omega_{i}^{AB} \left[ \left(1-x_p^{i/f} \right)x_{i} + \left(1-x_{i} \right)x_p^{i/f} \right] f_{i} \frac{z_{ip}}{2} \nonumber\\
    & {}+ \left[\delta_i^{AA}\left(1-x_{i} + 1-x_{p}^{i/f} \right) + \delta_i^{BB}\left(x_{i}+x_p^{i/f} \right)\right]f_{i} \frac{z_{ip}}{4}, \label{eq:dU_binary_d}\\[20pt]
    \Delta \bar{U}^{\,ref}_{b,i} =& 
\left[\left(1-x_b \right) \Delta \bar{U}_{ref(b)}^A + x_b\Delta \bar{U}_{ref(b)}^B \right] \left(1 - f_{i} - f_p \right) \nonumber \\
& {}+ \left[\left(1-x_{i}\right) \Delta \bar{U}_{\,ref(b)}^A + x_{i}\Delta \bar{U}_{\,ref(b)}^B \right]f_{i}, \label{eq:dU_binary_b}\\[20pt]
	\Delta \bar{U}_p =& \left[\frac{\bar{U}_f^{A_m B_n}}{m+n} - \left(1 - x_p \right) \bar{U}_{std}^A - x_p \bar{U}_{std}^B \right] f_p. \label{eq:dU_binary_c}
  \end{align}
\end{subequations}

\newpage
\subsubsection{Free Energy Function}

The molar free energy change for the formation of the binary system configuration,
$\Delta \bar{G}_{bin}$, is written in terms of the molar enthalpy change, $\Delta \bar{H}_{bin}$,
and the molar entropy change, $\Delta \bar{S}_{bin}$, as

\begin{equation} \label{eq:delg_ref1} \Delta \bar{G}_{bin} =
  \Delta \bar{H}_{bin} - T \Delta \bar{S}_{bin}.
\end{equation}

Neglecting any change in volume, $\Delta \bar{H}_{bin}$ can be
approximated to equal $\Delta \bar{U}_{bin}$ (Eq. \ref{eq:dU_binary}). $\Delta \bar{S}$ is taken as the change configurational
entropy, $\Delta \bar{S}^{\,mix}_{b}$ and $\Delta \bar{S}^{\,mix}_{i}$, associated with the the random
mixing involved in the formation of $b$ and $i$ regions
of the system, respectively, from the pure component reference states. $\Delta \bar{S}_{bin}$ is obtained from the Boltzmann's equation as an
additive expression involving region-size scaled entropy contributions.

\begin{subequations} \label{eq:dels_ref1}
	\begin{align}
    \Delta \bar{S}_{bin} =& \, \Delta \bar{S}^{\,mix}_b + \Delta \bar{S}^{\,mix}_i \tag{\ref{eq:dels_ref1}} \\[10pt]
	\Delta \bar{S}^{\,mix}_{b} =& -R \left[\left(1-x_b \right) \text{ln} \left(1-x_b \right) + x_b\,\text{ln}\,x_b \right] \left(1-f_i-f_p\right) \label{eq:dels_ref1a}\\[10pt]
    \Delta \bar{S}^{\,mix}_{i} =& - R \left[\left(1-x_i \right) \text{ln} \left(1-x_i\right) + x_i\,\text{ln}\,x_i \right]f_i
	\end{align}
\end{subequations}

\noindent In the above, the configurational entropy change associated with the formation of region $p$ is
neglected considering that both $A$ and $B$ atoms in $A_mB_n$ compound
occupy their corresponding sub-lattice sites. However, $\Delta \bar{S}$
terms for $p$, and also for the change in state from standard to
reference state for $b$ and $i$, are considered and coupled
appropriately to the related $\Delta \bar{U}$ terms to obtain the Gibbs free
energy function in Eq. (\ref{eq:dG_binary}). While
evaluating this function for an actual alloy system, free energy
values/expressions are taken from thermodynamic literature/database. These values/expressions are generally obtained form empirical measurements or first-principles calculations and inherently account for
configuration or vibrational entropy contributions, even though these are
ignored in the model itself.

\newpage
The molar free energy function for the binary system, $\Delta \bar{G}_{bin}$, is obtained from Eqs. (\ref{eq:dU_binary}), (\ref{eq:delg_ref1}) and (\ref{eq:dels_ref1}) as:

\begin{subequations} \label{eq:dG_binary}
  \begin{align} 
   \Delta \bar{G}_{bin} &= \Delta \bar{G}^{\,mix}_{b} + \Delta \bar{G}^{\,mix}_{i} + \Delta \bar{G}_{ib} + \Delta \bar{G}_{ip} + \Delta \bar{G}^{\,ref}_{b,i} + \Delta \bar{G}_p, \tag{\ref{eq:dG_binary}}\\[20pt]
    \text{where,} \nonumber\\[20pt]
    \Delta \bar{G}^{\,mix}_{b} &= \Delta \bar{U}^{\,mix}_{b} - T \Delta \bar{S}^{\,mix}_{b}, \label{eq:dG_binary_a}\\[20pt]
    \Delta \bar{G}^{\,mix}_{i} &= \Delta \bar{U}^{\,mix}_{i} - T \Delta \bar{S}^{\,mix}_{i}, \label{eq:dG_binary_b}\\[20pt]
    \Delta \bar{G}_{ib} &= \Delta \bar{U}_{ib}, \label{dG_bainry_c}\\[20pt]
    \Delta \bar{G}_{ip} &= \Delta \bar{U}_{ip}, \label{dG_bainry_d}\\[20pt]
     \Delta \bar{G}^{\,ref}_{b,i} &= \left[\left(1-x_b \right) \Delta \bar{G}_{ref(b)}^A + x_b\Delta \bar{G}_{ref(b)}^B \right] \left(1 - f_{i} - f_p \right) \nonumber\\
    & {}+ \left[\left(1-x_{i} \right) \Delta \bar{G}_{ref(b)}^A + x_{i}\Delta \bar{G}_{ref(b)}^B \right]f_{i}, \label{eq:dG_binary_b}\\[20pt]
	\Delta \bar{G}_p &= \Delta \bar{G}_f^{A_mB_n} f_p, \label{eq:dG_binary_c}\\[20pt]
    \Delta \bar{G}_f^{A_mB_n} &= \frac{\bar{G}_f^{A_m B_n}}{m+n} - \left(1 - x_p \right) \bar{G}_{std}^A - x_p \bar{G}_{std}^B. 
  \end{align}
\end{subequations}

\newpage
\subsection{Ternary Model}

As with the binary system, we consider the ternary system to consist of atomic regions $b$, $i$ and $p$, and bonding regions $b$, $ii$, $ib$ and $ip$. Additional terms and relations arise due to the presence of the ternary component $C$; these are presented below.

The internal energy of the system configuration and the initial configuration are given by:

\begin{align} \label{eq:bin_tern1}
    U_{sys} &= \sum\limits_{reg = b,i,p} \left(N_{reg}^A E_{reg}^A + N_{reg}^B E_{reg}^B + N_{reg}^C E_{reg}^C \right) \\
    U_{std} &= N_{total}^A E_{std}^A + N_{total}^B E_{std}^B + N_{total}^C E_{std}^C 
\end{align}

\noindent Since we consider $C$ to be insoluble in the $A_mB_n$ precipitate, the total number of $C$ atoms in the system, $N_{total}^C$, is given by:

\begin{align} \label{seq:NC_tern}
	N_{total}^C = N_b^C + N_i^C
\end{align}

\noindent Relations between the number of atoms of $A$ and the number of bonds connecting atoms of $A$ and specific to various bonding regions are obtained as:

\begin{align} \label{eq:bin_tern2}
    N_b^A z_b & = 2N_{ref(b)}^{AA} = 2N_b^{AA} + N_b^{AB} + N_b^{AC}\\
    N_i^A z_i & = 2N_{ref(i)}^{AA} = 2N_i^{AA} + N_i^{AB} + N_i^{AC} \nonumber\\
    & = \left(2N_{ii}^{AA} + N_{ii}^{AB} + N_{ii}^{AC} \right) +
    \left(2N_{ib}^{AA} + N_{ib}^{AB} + N_{ib}^{AC} \right) \nonumber\\ 
    &+ \left(2N_{ip}^{AA} + N_{ip}^{AB} + N_{ip}^{AC} \right)
\end{align}

\noindent Similar relations are obtained for components $B$ and $C$. Relations for $N^{bonds}_r$ and $P^{kl}_r$ for the ternary system are
listed in the Table \ref{table:ternary} in the main paper. $N_b^k$ and $N_i^k$ can be expressed in terms of region-specific volume fractions and concentrations and total number of atoms as given below; $y_b$ and $y_i$ are the concentrations of $C$ in $b$ and $i$, respectively.

\begin{align} \label{seq:tern_atom}
\begin{aligned}
    N_b^A &= (1-f_i-f_p) N_{\circ} \left(1-x_b - y_b\right), \\ 
    N_b^B &= (1-f_i-f_p) N_{\circ} \, x_b, \\
    N_b^C &= (1-f_i-f_p) N_{\circ} \, y_b,
\end{aligned}
&&
\begin{aligned}
    N_i^A &= f_i N_{\circ} \left(1-x_i - y_i\right) \\ 
    N_i^B &= f_i N_{\circ} \, x_i, \\
    N_i^C &= f_i N_{\circ} \, y_i.
\end{aligned}
\end{align}

\noindent In addition to the mass balance relation for solute $B$, which is given by Eq. (\ref{eq:cons_x}), the mass balance relation for solute $C$ is obtained as:

\begin{align} \label{eq:constraint_tern}
    y_{\circ} = y_b \left(1-f_i-f_p \right) + y_i f_i
\end{align}

\noindent As with $\omega_r^{AB}$ (Eq. \ref{eq:omega_def}), regular solution parameters for mixing of $B$ and $C$ and mixing of $A$ and $C$ are defined with energies characteristic of $r$ ($b$ or $i$) as:

\begin{align} \label{seq:omega_tern}
	\omega_r^{AC} = E_r^{AC} - \frac{E_r^{AA}+E_r^{CC}}{2}, \qquad \omega_r^{BC} = E_r^{BC} - \frac{E_r^{BB}+E_r^{CC}}{2}
\end{align}

\noindent The interface energy penalty parameter for $CC$ bonds at the interface or transition regions is defined by:

\begin{align} \label{seq:diC}
	\delta_i^{CC} = \frac{2}{z_i}\left(E_{ref(i)}^C - E_{ref(b)}^C\right)
\end{align}

\noindent The energy difference for $C$ between its pure component reference state characteristic of $b$ and its pure component standard state is:

\begin{align} \label{seq:refU_tern}
	\Delta U_{std \rightarrow ref(b)}^C = E_{ref(b)}^C - E_{std}^C
\end{align}

\newpage
\subsubsection{Internal Energy Function}

The internal energy function for the ternary system, $\Delta U_{tern}$, is now obtained using the modified and the additional relations for the ternary system and following the derivation presented for the binary model.

\begin{align} \label{seq:dU_tern_sys}
	\Delta \bar{U}_{tern} = \Delta \bar{U}^{\,mix}_b + \Delta \bar{U}^{\,mix}_{i} + \Delta \bar{U}_{ib} + \Delta \bar{U}_{ip} + \Delta \bar{U}^{\,ref}_{b,i} + \Delta \bar{U}_p
\end{align}

\noindent where,

\begin{align} \label{seq:dU_tern_bmix}
  \begin{split}
	\Delta \bar{U}^{\,mix}_b =& \left[\omega_b^{AB} \left(1 - x_b - y_b \right) x_b + \omega_b^{BC} x_b y_b + \omega_b^{AC} \left(1 - x_b - y_b \right) y_b \right] \left(1 - f_{i} - f_p \right) z_b
    \end{split}
\end{align}

\begin{align} \label{seq:dU_tern_imix}
  \begin{split}
	\Delta \bar{U}^{\,mix}_{i} =& \left[\omega_{i}^{AB} \left(1 - x_{i} - y_{i} \right) x_{i} + \omega_{i}^{BC} x_{i} y_{i} + \omega_{i}^{AC} \left(1 - x_{i} - y_{i} \right) y_{i} \right]  f_{i} z_{ii}\\
    &+ \left[\delta_i^{AA}\left(1-x_{i}-y_{i}\right) + \delta_i^{BB}x_{i} + \delta_i^{CC}y_{i} \right]f_{i} \frac{z_{ii}}{2}
    \end{split}
\end{align}

\begin{align} \label{seq:dU_tern_ib}
  \begin{split}
	\Delta \bar{U}_{ib} =& \left\{\omega_{i}^{AB} \left[ \left(1-x_b-y_b \right)x_{i} + \left(1-x_{i}-y_{i} \right)x_b \right] + \omega_{i}^{BC} \left(x_b y_{i} + x_{i} y_b \right) \right. \\
    &\qquad {}+ \left. \omega_{i}^{AC} \left[\left(1-x_b-y_b\right)y_{i} + \left(1-x_{i}-y_{i} \right)y_b \right] \right\} f_{i} \frac{z_{ib}}{2} \\
    &+ \left[\delta_i^{AA}\left(1-x_{i}-y_{i} + 1-x_{b}-y_{b} \right) + \delta_i^{BB}\left(x_{i}+x_b \right) + \delta_i^{CC}\left(y_{i}+y_b \right) \right]f_{i} \frac{z_{ib}}{4} \\
    \end{split}
\end{align}

\begin{align} \label{seq:dU_tern_ip}
  \begin{split}
	\Delta \bar{U}_{ip} =& \left\{\omega_{i}^{AB} \left[ \left(1-x_p^{i/f} \right)x_{i} + \left(1-x_{i}-y_{i} \right)x_p^{i/f} \right] + \omega_{i}^{BC} x_p^{i/f} y_{i} \right. {}\\
    &\qquad {}+ \left. \omega_{i}^{AC} \left(1-x_p^{i/f} \right)y_{i} \right\} f_{i} \frac{z_{ip}}{2} \\
    &+ \left[\delta_i^{AA}\left(1-x_{i}-y_{i} + 1-x_{p}^{i/f} \right) + \delta_i^{BB}\left(x_{i}+x_{p}^{i/f} \right) + \delta_i^{CC} y_{i} \right]f_{i} \frac{z_{ip}}{4} \\
    \end{split}
\end{align}

\begin{align} \label{seq:dU_tern_ref}
  \begin{split}
	\Delta \bar{U}^{\,ref}_{b,i} =& \left[\left(1-x_b-y_b\right) \Delta U_{ref(b)}^A + x_b\Delta U_{ref(b)}^B + y_b\Delta U_{ref(b)}^C \right] \left(1 - f_{i} - f_p \right) \\
    &+ \left[\left(1-x_{i}-y_{i}\right) \Delta U_{ref(b)}^A + x_{i}\Delta U_{ref(b)}^B + y_{i}\Delta U_{ref(b)}^C \right]f_{i}
    \end{split}
\end{align}

\begin{align} \label{seq:dU_tern_p}
  \begin{split}
	\Delta \bar{U}_{p} =& \left[\frac{U_f^{A_m B_n}}{m+n} - \left(1 - x_p \right) U_{std}^A - x_p U_{std}^B \right] f_p 
    \end{split}
\end{align}

\newpage
\subsubsection{Free Energy Function}

The configuration entropy for the ternary system, $\Delta \bar{S}_{tern}$ is obtained as:
\begin{subequations} \label{eq:dels_ref2}
	\begin{align}
    \Delta \bar{S}_{tern} =& \, \Delta \bar{S}^{\,mix}_{b} + \Delta \bar{S}^{\,mix}_i \tag{\ref{eq:dels_ref2}} \\[10pt]
    \text{where,} \nonumber \\[10pt]
	\Delta \bar{S}^{\,mix}_{b} =& -R \left[\left(1-x_b-y_b\right) \text{ln} \left(1-x_b-y_b\right) + x_b\,\text{ln}\,x_b + y_b\,\text{ln}\,y_b \right] \left(1-f_i-f_p\right) \label{eq:dels_ref2a}\\[10pt]
    \Delta \bar{S}^{\,mix}_{i} =& - R \left[\left(1-x_i-y_i \right) \text{ln}
      \left(1-x_i-y_i\right) + x_i\,\text{ln}\,x_i +
      y_i\,\text{ln}\,y_i \right] f_i
	\end{align}
\end{subequations}

\vspace{5mm}
The free energy change for the formation of the ternary system configuration, $\Delta \bar{G}_{tern}$, is given by:
\begin{align} \label{seq:G_tern_def}
	\Delta \bar{G}_{tern} = \Delta \bar{U}_{tern} - T \Delta \bar{S}_{tern}
\end{align}

\noindent Using Eqs. (\ref{seq:dU_tern_sys}), (\ref{eq:dels_ref2}) and (\ref{seq:G_tern_def}), and coupling $\Delta \bar{U}$ terms in Eqs. (\ref{seq:dU_tern_ref}) and (\ref{seq:dU_tern_p}) with corresponding $-T\Delta \bar{S}$ terms, the final expression of $\Delta \bar{G}_{tern}$ is obtained as:

\begin{align}
	\Delta \bar{G}_{tern} = \Delta \bar{G}^{\,mix}_b + \Delta \bar{G}^{\,mix}_{i} + \Delta \bar{G}_{ib} + \Delta \bar{G}_{ip} + \Delta \bar{G}^{\,ref}_{b,i} + \Delta \bar{G}_p
\end{align}
where,
\begin{align}
  \begin{split}
	\Delta \bar{G}^{\,mix}_b =& \Delta \bar{U}^{\,mix}_b + \Delta \bar{S}^{\,mix}_i \\
    \end{split}
\end{align}

\begin{align}
  \begin{split}
	\Delta \bar{G}^{\,mix}_{i} =& \Delta \bar{U}^{\,mix}_{i} + \Delta \bar{S}^{\,mix}_{i}\\
    \end{split}
\end{align}

\begin{align}
  \begin{split}
	\Delta \bar{G}_{ib} =& \Delta \bar{U}_{ib}\\
    \end{split}
\end{align}

\begin{align}
  \begin{split}
	\Delta \bar{G}_{ip} =& \Delta \bar{U}_{ip} \\
    \end{split}
\end{align}

\begin{align}
  \begin{split}
	\Delta \bar{G}^{\,ref}_{b,i} =& \left[\left(1-x_b-y_b\right) \Delta G_{ref(b)}^A + x_b\Delta G_{ref(b)}^B + y_b\Delta G_{ref(b)}^C \right] \left(1 - f_{i} - f_p \right) \\
    &+ \left[\left(1-x_{i}-y_{i}\right) \Delta G_{ref(b)}^A + x_{i}\Delta G_{ref(b)}^B + y_{i}\Delta G_{ref(b)}^C \right]f_{i}
    \end{split}
\end{align}

\begin{align}
  \begin{split}
	\Delta \bar{G}_{p} =& \Delta \bar{G}_f^{A_m B_n} f_p 
    \end{split}
\end{align}

\begin{align}
  \begin{split}
	\Delta \bar{G}_{f}^{A_mB_n} =& \left[\frac{G_f^{A_m B_n}}{m+n} - \left(1 - x_p \right) G_{std}^A - x_p G_{std}^B \right] f_p 
    \end{split}
\end{align}

\clearpage

\input{Thermo_data.tex}

\clearpage

\putbib
\end{bibunit}

\end{widetext}



%% file: Thermo_data.tex
\section{Thermodynamic data} \label{SI2:therm_data}

Input to the thermodynamic parameters for the ternary model are obtained for Mg-Sn-Zn system from thermodynamic databases and literature. These are presented in this section in units of kJ/mol. Mg, Sn and Zn correspond to A, B and C, respectively.

For hcp-Mg solid-solution phase, binary interaction parameters ($L_0^{kl}$) of the Redlich-Kister-Muggianu free energy expression, describing the excess free energy contribution from non-ideal interactions, were evaluated by Meng et. al. \cite{meng2010thermodynamic}. While higher order terms corresponding to ternary interactions were set to zero in their model, interaction parameters ($L_1^{kl}$) corresponding to composition dependent terms were included in the model and evaluated. However, $L_1^{kl}$ parameters, describing the sub-regular solution behavior, are ignored in our study for simplicity, and thus only $L_0^{kl}$ terms are considered. The temperature dependent $L_0^{kl}$ obtained from \cite{meng2010thermodynamic} are:

\begin{equation}
    \begin{aligned}
        & L_0^{AB} = (-26256.5 + 6.234 \, T) \times 10^{-3}, \\
        & L_0^{BC} = 30453 \times 10^{-3}, \\
        & L_0^{AC} = (-3056.82 + 5.63801 \, T) \times 10^{-3}. \\
    \end{aligned}
\end{equation}
\\
\noindent The regular-solution interaction parameters ($\omega_b^{kl}$) of the bulk region modeled in the present study are related to $L_0^{kl}$ through:

\begin{align}
    & \omega_b^{AB} = \frac{L_0^{AB}}{z_b}, \quad \omega_b^{BC} = \frac{L_0^{BC}}{z_b}, \quad \omega_b^{AC} = \frac{L_0^{AC}}{z_b}.
\end{align}
\\
\noindent The free energy of formation of Mg$_2$Sn, denoted by $ G^{A_m B_n}_f$ in the present model, was determined in \cite{meng2010thermodynamic} from experimental heat capacity and heat content measurements reported in literature,

\begin{equation}
    \begin{aligned}
        & G^{A_mB_n}_f = (-96165.9 + 339.999 \, T - 66.285 \, T \,\text{ln}\,(T) - 0.0121662 \, T^2 \\
        & \qquad \qquad + 96000 \, T^{-1} + 3.33828 \times 10^{-7} \, T^3) \times 10^{-3}. \\
    \end{aligned}
\end{equation}
\\
\noindent The standard state energies, $G_{std}^k$, and the difference in free energies between the standard state and the reference state, $\Delta G^k_{ref(b)}$, of the present model are obtained from the SGTE thermodynamic data compiled by Dinsdale \cite{dinsdale1991sgte}. These respective quantities are defined in \cite{dinsdale1991sgte} as the Gibbs energy evaluated with respect to enthalpy of the “Standard Element Reference” (which is the reference phase at 298.15 K for $k$), and the difference in Gibbs energy between the pertinent phase of $k$ and its reference phase. The expressions for Mg, Sn and Zn are given below:

\begin{align}
    \begin{split}
        & G_{std}^A = (-8367.34 + 143.675547 \, T - 26.1849782 \, T \,\text{ln}\,(T) + 0.4858 \times 10^{-3} \, T^2 ,\\
        & \qquad \qquad - 1.393669 \times 10^{-6} \, T^3 + 78950 \, T^{-1}) \times 10^{-3} ,\\
        & G_{std}^B = (-5855.135 + 65.443315 \, T - 15.961 \, T \,\text{ln}\,(T) - 18.8702 \times 10^{-3} \, T^2 ,\\
        & \qquad \qquad + 3.121167 \times 10^{-6} \, T^3 - 61960 \,T^{-1}) \times 10^{-3} ,\\
        & G_{std}^C = (-7285.787 + 118.470069 \, T - 23.701314 \, T \,\text{ln}\,(T) - 1.712034 \times 10^{-3} \, T^2 ,\\
        & \qquad \qquad - 1.264963 \times 10^{-6} \, T^3) \times 10^{-3} ,
    \end{split}
\end{align}
\begin{align}
    \begin{split}
        & \Delta G_{ref(b)}^A = 0 ,\\
        & \Delta G_{ref(b)}^B = (3900 - 4.4 \, T) \times 10^{-3} ,\\      
        & \Delta G_{ref(b)}^C = 0.
    \end{split}
\end{align}
\\
\section{Interface energy penalty} \label{SI3:delta}
The interface bond energy parameters, $\delta_i^{AA}$, representing the energy of interface-type $AA$ bond relative to the bulk-type $AA$ bond can be related to the free energy of the interface, $\gamma^A_i$, between $A$-rich solid-solution and $A_mB_n$ precipitate through,

\begin{align}
    \delta_i^{AA} = \frac{2}{z_i} \sigma \gamma^A_i.
\end{align}

\noindent Here, $\sigma$ is the molar surface area of the interface, which is given by $N_{avg}\Omega^{2/3}$, and $\Omega$ is the atomic volume of A. For an incoherent heterophase interface between Mg-rich solid-solution and Mg$_2$Sn precipitate, Katsman et. al. \cite{katsman2007modeling} reported an average value of $410$ kJ/mol, which was estimated using a Langer-Schwartz model for precipitation and hardening, and an experimental measurement of aging response. Using this, $\delta_i^{AA}$ of $3.5$ kJ/mol is obtained as a rough input for the present model.